\documentclass[11pt]{article}
\usepackage{setspace}
\usepackage{latexsym,amsthm,amsmath,amssymb,url}
\usepackage{array}
\usepackage{enumitem}
\usepackage{algorithm}
\usepackage{algorithmic}
\usepackage[round]{natbib}
\usepackage{eurosym}
\usepackage{subcaption}
\usepackage{pgfplots}

\definecolor{darkgreen}{RGB}{0, 150, 0}
\usepackage[defaultcolor=darkgreen]{changes}

\usepackage{graphics}
\usepackage{color}
\usepackage{colortbl}
\usepackage{booktabs}
\usepackage{tikz}
\usetikzlibrary{fit}
\usepackage{tikzsymbols}

\theoremstyle{definition}
\newtheorem{lemma}{Lemma}

\theoremstyle{definition}
\newtheorem{definition}{Definition}
\theoremstyle{definition}
\newtheorem{example}{Example}

\theoremstyle{definition}
\newtheorem*{theorem*}{Theorem}
\theoremstyle{definition}
\newtheorem{theorem}{Theorem}
\theoremstyle{definition}
\newtheorem{observation}{Observation}

\newtheorem*{algo*}{Algorithm}

\newtheorem{innercustomthm}{Theorem}
\newenvironment{customthm}[1]
  {\renewcommand\theinnercustomthm{#1}\innercustomthm}
  {\endinnercustomthm}

\newcommand{\2}{\vspace{0.2cm}}

\newcommand{\red}[1]{{\color{red}#1}}
\newcommand{\blue}[1]{{\color{blue}#1}}
\newcommand{\green}[1]{{\color{greengray}#1}}

\newcommand{\gray}[1]{{\color{gray}#1}}

\newcommand{\set}[1]{\left\{#1\right\}}
\newcommand{\Set}[1]{\Bigl\{#1\Bigr\}}

\newcommand{\Men}{M}
\newcommand{\man}[1]{m_{#1}}
\newcommand{\Wom}{W}
\newcommand{\wom}[1]{w_{#1}}

\newcommand{\pref}[1]{\ensuremath{\succ_{#1}}}

\newcommand{\match}{\mu}

\newcommand{\rds}[1]{%
  \tikz[baseline={(textnode.base)}] \node[inner sep=0pt, outer sep=0pt] (textnode) {$#1$};%
  \tikz[overlay] \node[red, draw, thick, rectangle, inner sep=2pt, outer sep=0pt, fit={(textnode)}, shift={(-7pt, 2pt)}] {};%
}
\newcommand{\blc}[1]{%
  \tikz[baseline={(textnode.base)}] \node[inner sep=0pt, outer sep=0pt] (textnode) {$#1$};%
  \tikz[overlay] \node[blue, draw, thick, circle, inner sep=0.2pt, outer sep=0pt, fit={(textnode)}, shift={(-7pt, 2pt)}] {};%
}

 \newcommand{\WithoutColor}[1]{}       
\newcommand{\matchingC}[1]{yellow}

\usepackage[pdftex, citecolor=blue, colorlinks]{hyperref}

\newcolumntype{L}[1]{>{\raggedright\let\newline\\\arraybackslash\hspace{0pt}}m{#1}}
\newcolumntype{C}[1]{>{\centering\arraybackslash\vspace*{\fill}\vspace{-0.5ex}}m{#1}<{\vspace*{\fill}}}
\newcolumntype{R}[1]{>{\raggedleft\let\newline\\\arraybackslash\hspace{0pt}}m{#1}}

\definecolor{lightgray}{RGB}{229, 228, 226}
\definecolor{greengray}{RGB}{163, 201, 170}

\pgfplotsset{compat=1.13}
\usepgfplotslibrary{groupplots}

\begin{document}

\title{Speeding up deferred acceptance\thanks{We thank Frances Ruane and Joel Sobel for comments.}}
\author{
Gregory Z. Gutin\footnote{Computer Science Department, Royal Holloway University of London.} \textsuperscript{,}\footnote{School of Mathematical Sciences and LPMC, Nankai University, Tianjin, China.} \and Daniel Karapetyan\footnote{School of Computer Science, University of Nottingham.} \and Philip R.\ Neary\footnote{Economics Department, Royal Holloway University of London.} \and Alexander Vickery\footnote{BDO LLP.} \and Anders Yeo\footnote{IMADA, University of Southern Denmark.} \textsuperscript{,} \footnote{Department of Mathematics, University of Johannesburg.}
}

\date{\today}

\maketitle

\begin{abstract}
  \noindent
  A run of the deferred acceptance (DA) algorithm may contain proposals that are sure to be rejected.
  We introduce the \emph{accelerated deferred acceptance} algorithm that proceeds in a similar manner to DA but with sure-to-be rejected proposals ruled out.
Accelerated deferred acceptance outputs the same stable matching as DA but does so more efficiently:\ it terminates in weakly fewer rounds, requires weakly fewer proposals, and final pairs match no later.
  Computational experiments show that these efficiency savings can be strict.
\end{abstract}

\newpage

\section{Introduction}\label{sec:intro}

In this paper we introduce a new algorithm to find a stable matching in two-sided matching markets.
Our \emph{accelerated deferred acceptance} algorithm combines the classic deferred acceptance (DA) algorithm of \cite{GaleShapley:1962:AMM} with insights from the iterated deletion of unattractive alternatives (IDUA) procedure \citep{BalinskiRatier:1997:,GutinNeary:2023:GEB}.
Accelerated deferred acceptance is reminiscent of DA since it is based on sequential proposals, rejections, and tentative acceptances.
Accelerated deferred acceptance borrows from IDUA by truncating preference lists so that future, sure-to-be rejected proposals are prevented from taking place.

The two-sided markets that we consider have ``men'' on one side and ``women'' on the other.
With men in the role of proposers, accelerated deferred acceptance (hereafter ADA) diverges from DA in just one way:\ once a woman has a proposal, she rejects all men ranked below her top proposer and not just those that proposed at the same time.\footnote{Operationally, the difference is clear. Semantically, however, the change is minor. In fact, ADA deviates from DA (see Algorithm~\ref{algo:DA} for the formal statement) in just one term:\ the word ``men'' -- typeset in \textbf{\red{red}} for emphasis -- replaces the word ``proposers''.}

\begin{algorithm}
  \renewcommand{\thealgorithm}{}
  \caption{Accelerated deferred acceptance}\label{algo:ADA}\,
  
  Initialise all men as single.
  Every round of the algorithm proceeds as follows:

  \medskip
  \noindent
  1. Each single man proposes to his most preferred woman who has not yet rejected him.\\
  2. Each woman with at least one proposal tentatively accepts her top proposer and rejects all \textbf{\red{men}} that she ranks below him.\\
  3. Any man who is not matched becomes single.

  \medskip
  \noindent
  When there is a round in which no man is rejected, return the current pairs.
\end{algorithm}

Theorems~\ref{thm:ADAstable} and \ref{thm:ADAstrategyProof} confirm that ADA is similar to DA in two ways:\ ADA produces the same output and is strategy proof for the proposing side.
However, the \emph{pre-emptive rejections} allowed by ADA have consequences for both sides.
Once a woman is proposed to, she will never again receive a proposal from a man that she ranks below her current partner.
As such, whenever a new proposal arrives, the woman is guaranteed to trade up.
For men the situation is also different.
A man may receive a rejection out of the blue.
Such pre-emptive rejections may be unexpected at the time, but, in the event that the man becomes single at some point in the future, can be informative when deciding who he ought propose to next.

ADA arrives at the man-optimal stable matching via a different path to that taken by DA.
This alternate path is more efficient according to certain measures:\ number of proposals required for the market to clear; number of rounds for the market to clear; round in which final pairs first match.\footnote{We regard these three measures as objective because they can be applied widely to all two-sided market clearing procedures.}
When benchmarked against DA, ADA requires weakly fewer proposals (Theorem~\ref{thm:ADAproposals}), takes weakly fewer rounds (Theorem~\ref{thm:ADArounds}), and final pairs are matched no later (Theorem~\ref{thm:ADAearlier}).
However, these come at a cost because ADA does requires weakly more total rejections (defined as direct rejections and pre-emptive rejections combined).

Theorems~\ref{thm:ADAproposals}-\ref{thm:ADAearlier} guarantee no efficiency losses but are silent on the extent of the efficiency gains.
We explore these via computational experiments on simulated data.\footnote{Our simulations were run in Python. All code is publicly available at XXXX.}
This is a non-trivial task because the number of instances of a two-sided matching problem of size $n$ is $(n!)^{2n}$, so even enumerating all of them is computationally infeasible for $n=5$ and up.
For this reason, we propose a novel instance generator that allows one to sample random instances for any market size.
A scalar parameter, $c\in [0,1]$, biases the sampling distribution.
The value $c=0$ corresponds to preferences drawn uniformly from the set of all preferences.
The value $c=1$ means that all individuals on the same side of the market have identical preferences, a property known as a ``universal ranking'' \citep{HolzmanSamet:2014:GEB}.\footnote{While statistically unlikely, a universal ranking is a common assumption in marriage markets with endowments \citep{ColeMailath:1992:JPE,BurdettColes:1997:QJE,Eeckhout:1999:IER}. A universal ranking for one side is hardwired in to certain third-level admissions systems. An example is the centralised system in Ireland, the ``Leaving Cert", where marks out of 625 in state-administered exams are the sole determinant of university entry.}

\begin{figure}[!ht]
  \centering
  \includegraphics[width=0.94\textwidth,trim=0.5cm 0.6cm 0.5cm 1cm,clip]{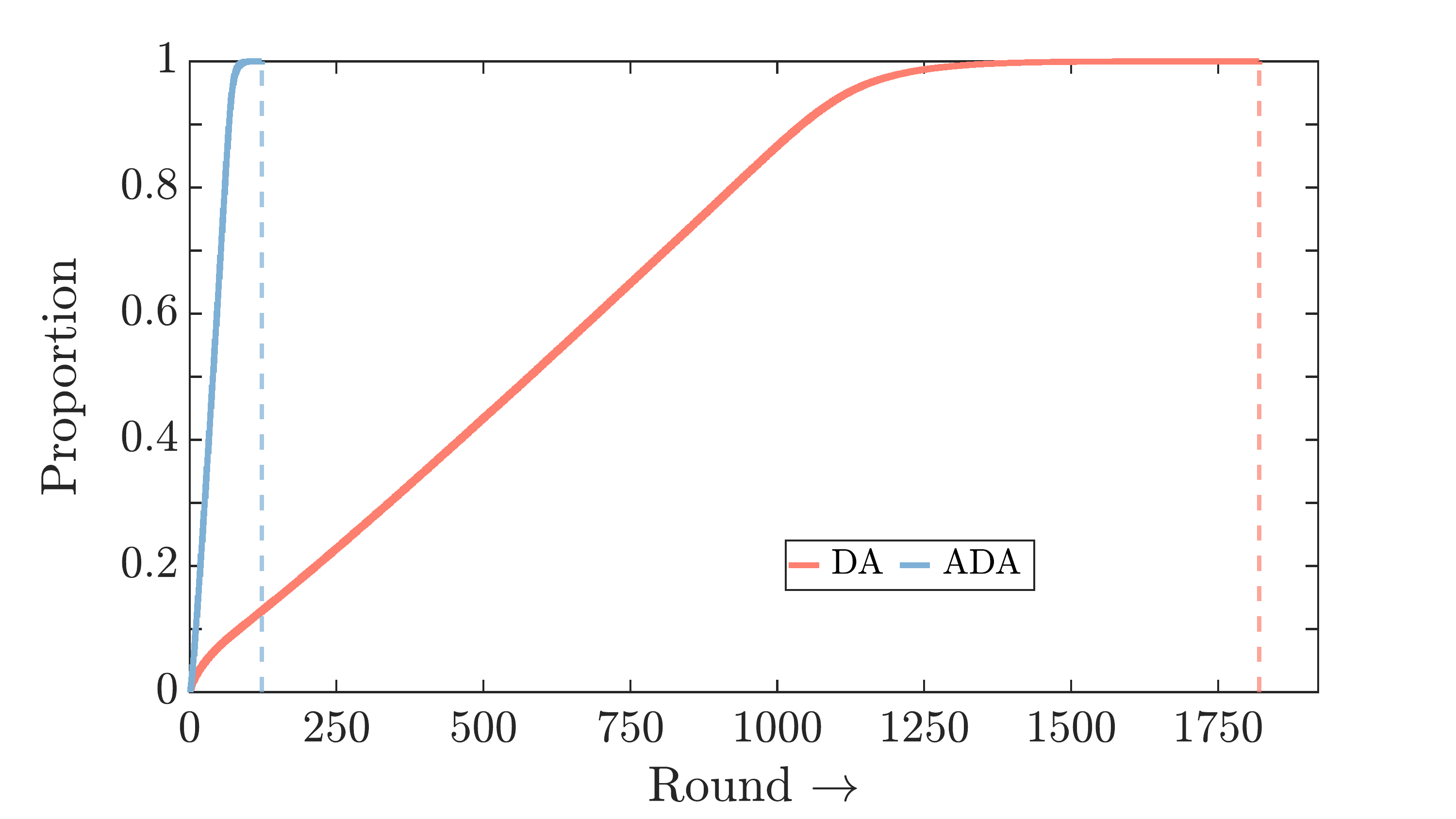}
  \caption{The proportion of final pairs matched by round.}
  \label{fig:probability}
\end{figure}

Our simulations show that ADA's efficiency savings can be substantial.\footnote{All results are averaged over 1,000 draws.}
For example, in markets with 4,096 $(=2^{12})$ participants on each side and $c=0.9$, the average number of proposals used by DA is over 7,500,000.
In contrast, ADA required only 208,000.
This is a reduction of 96\%.
As regards the round in which final pairs first match, again ADA fares better.
Figure~\ref{fig:probability} plots this metric, with ADA in \blue{blue} and DA in \red{red}, for a market with 1,024 $(=2^{10})$ participants and $c = 0.9$.
While ADA terminates after 123 rounds, at this point, DA is only 7\% of the way to completion as it takes 1,820 rounds.
Moreover, as of round 123, DA has only identified 13\% of pairs that will end up together; the remaining 87\% of individuals are either unmatched or are currently matched with someone who they will not end up with.

We hope that the efficiencies of ADA over DA will be of general interest since variations of DA appear in labour markets \citep{CrawfordKnoer:1981:E,KelsoCrawford:1982:E}, spectrum auctions \citep{MilgromSegal:2020:JPE}, and school choice \citep{AbdulkadirogluSonmez:2003:AER}.
In each of these environments, all three of our metrics appear relevant.
Consider, for example, the savings from ruling out sure-to-be rejected proposals.
An attempt to recruit a worker with a contract inferior to what they currently hold is unlikely to work.
In an auction, why submit a bid below the ask?
Filling out an application form for a school that is out of reach is nothing but a tax on the time of both parties.
We refer to a round in which all proposals are rejected as an \emph{idle round}.
ADA never produces an idle round.
DA can contain an idle round and in fact it can have many.

We conclude the introduction with a brief discussion of interpretation.
One can understand ADA in a similar light to DA -- a setting where men traverse a dance floor to ask women for a dance -- but with the algorithm clever enough that it can identify irrelevant proposals coming down the tracks and barring them.
Other readings are possible.
One way to interpret the difference in the two procedures is in the level of informational feedback.\footnote{The role that information plays in markets cannot be understated -- \cite{Hayek:1945:AER} is the classic reference -- and matching markets are no exception. How feedback occurs seems particularly relevant for dynamic environments \citep{AkbarpourLi:2020:JPE,Doval:2022:TE}.}
Under DA, men only receive feedback in response to a rejected proposal.
And any time that a man does receive feedback, it is:\ ``your proposee rejected you''.
With ADA, the information flow is richer.
Feedback can come in two forms:\ ``your proposee rejected you'' and ``the following women are out of range''.
The former occurs only in response to a rejected proposal but the latter can arrive irrespective of the man's current status.
An increase in the feedback provided is beneficial if proposers use it.

Related to information feedback is strategic sophistication.
\citeauthor{GaleShapley:1962:AMM}'s DA asks very little of market participants:\ men blindly propose to their top available choice and all that is required of women is that they can rank proposers.\footnote{Note that men need not know the womens' preferences. In fact, arguably women need not be aware of their own preferences in advance; a woman simply needs to be able to identify the best candidate from arbitrary collections of men when it is necessary to do so, i.e., when more than one approaches.}
Another interpretation of ADA is that participants display sophisticated reasoning.\footnote{Blindly proposing to one's top available choice has the flavour of a greedy algorithm, and in the language of models of strategic sophistication \citep{Nagel:1995:AER,StahlWilson:1994:JEBO,StahlWilson:1995:GEB} might best be described as level-0 behaviour.}
In the first round each man proposes to his ideal partner; this seems reasonable since by definition she is unattached.
But in later rounds, might a strategic proposer contemplate making a proposal to a woman who already has a partner?
If the woman's current match trumps the would-be proposer, then proposing is not worth it.\footnote{\cite{BoHakimov:2019:EJ} report on an experimental test of the \emph{iterated deferred acceptance algorithm} of \cite{BoHakimov:2022:GEB} -- a dynamic mechanism where proposers make one proposal each round. \citeauthor{BoHakimov:2019:EJ} identify subjects who, from round 2 onwards, avoid proposals that won't lead anywhere, behaviour they refer as a ``skipping strategy''.}
Similarly, if dealing with unnecessary proposals is costly to a woman, it may be in her interest to convey this information to those who might tax her time in future.

\citeauthor{GaleShapley:1962:AMM}'s DA makes the standard algorithmic assumption that all operations (in this case, rejections) have unit cost. 
Accelerated deferred acceptance has two distinct kind of rejections, direct and pre-emptive, and we see no reason why the cost of a direct rejection and the cost of communicating a pre-emptive rejection must be equal.
It seems plausible to us that these costs could vary with market structure.
And depending on the relative costs, ADA may or may not be preferred to DA.

\subsection*{Related literature}\label{ssec:related}

In our opinion the main contribution of this paper is a natural augmentation of \citeauthor{GaleShapley:1962:AMM}'s original \citeyear{GaleShapley:1962:AMM} DA algorithm.
Given this, we limit our references to two comprehensive sources.
\cite{Roth:2008:IJGT} is a survey article devoted entirely to DA.
\citeauthor{Roth:2008:IJGT} documents the algorithm's importance and provides historical context.
The second reference is the \citeyear{EcheniqueImmorlica:2023:}  monograph ``Online and Matching-Based Market Design'' (\citeauthor{EcheniqueImmorlica:2023:}) that contains 32 chapters written by leading experts.
Discussion of DA appears throughout and plays a prominent role in Chapters 1, 7, 8, 10, 11, 12, 14, 15, and 24.\footnote{Direct reference to ``deferred acceptance'' or ``DA'' occurs in excess of 300 times.}

\subsection*{Structure of the paper}\label{ssec:structure}

The balance of the paper is as follows.
Section~\ref{sec:environment} fixes notation and reminds the reader of the environment, the DA algorithm, and the IDUA procedure.
In Section~\ref{sec:algorithm} we introduce the accelerated deferred acceptance algorithm and present its theoretical properties.
Section~\ref{sec:simulations} describes our instance generator and presents the results of our simulations.
Section~\ref{sec:conclusion} concludes and discusses potential avenues for future work.


\section{The environment}\label{sec:environment}

In Section~\ref{ssec:environment} we define the environment under study.
(It is precisely the one-to-one two-sided matching framework introduced by \citeauthor{GaleShapley:1962:AMM}.)
In Section~\ref{ssec:DA+IDUA} we recap the original \emph{deferred acceptance} (DA) algorithm and the \emph{iterated deletion of unattractive alternatives} (IDUA) procedure of \cite{BalinskiRatier:1997:} and \cite{GutinNeary:2023:GEB}.

\subsection{Matching problems}\label{ssec:environment}

Let $\Men$ be a set of $n$ men and let $\Wom$ be a set of $n$ women, where $n$ is a positive integer greater than or equal to 2.
Each man $\man{} \in \Men$ has a strict preference relation, $\pref{\man{}}$, over the set of all women, and each woman $\wom{} \in \Wom$ has a strict preference relation, $\pref{\wom{}}$, over the set of all men.
When man $\man{}$ prefers woman $\wom{}'$ to woman $\wom{}''$, we write $\wom{}' \pref{\man{}} \wom{}''$, with an analogous statement for the preferences of women.
(Occasionally it will be more convenient to present preferences as linearly ordered lists where the first entry on the list is the most preferred, and so on.
Which notation is being employed will be clear from context.)

\begin{definition}\label{def:instance}
  An instance of the stable matching problem, $P$, is defined as the pair $(\set{\pref{\man{}}}_{\man{}\in \Men}, \set{\pref{\wom{}}}_{\wom{}\in \Wom})$, where $\set{\pref{\man{}}}_{\man{}\in \Men}$ and $\set{\pref{\wom{}}}_{\wom{}\in \Wom}$ are the collections of preferences, one for each man and one for each woman.
\end{definition}

A \emph{matching} in $P$ is a mapping $\match$ from $\Men \cup \Wom$ to itself such that:\ for every man $\man{} \in \Men$, $\match(\man{}) \in \Wom$; for every woman $\wom{} \in \Wom$, $\match(\wom{}) \in \Men$; and for every man-woman pair $(\man{}, \wom{}) \in \Men \times \Wom$, $\match(\man{}) = \wom{}$ if and only if $\match(\wom{}) = \man{}$.

The following is the key definition proposed by \cite{GaleShapley:1962:AMM}.

\begin{definition}\label{def:blocking}
  Fix a matching $\match$ in $P$.
  Man $\man{}$ and woman $\wom{}$ form a \emph{blocking pair} with respect to $\match$ if $\wom{} \pref{\man{}} \match(\man{})$ and $\man{} \pref{\wom{}} \match(\wom{})$.
\end{definition}

In words, a man-woman pair $(\man{}, \wom{})$ is blocking with respect to matching $\match$ if both $\man{}$ and $\wom{}$ prefer each other over their partners in $\match$.
\citeauthor{GaleShapley:1962:AMM} defined stability by the absence of a blocking pair.

\begin{definition}\label{def:stableMatch}
  A matching $\match$ in $P$ with no blocking pairs is a \textit{stable matching}.
\end{definition}

\citeauthor{GaleShapley:1962:AMM} proved the following remarkable result.

\begin{theorem*}[\cite{GaleShapley:1962:AMM}]\label{thm:GaleShapley}
  Every instance of the stable matching problem possesses at least one stable matching.
\end{theorem*}

\subsection{DA and IDUA}\label{ssec:DA+IDUA}

In this section we define the DA algorithm and then the IDUA procedure.
We document both since ADA is a blend of the two.

Before beginning, one key difference between the two procedures is worth highlighting.
The DA algorithm returns a stable matching.
The IDUA procedure, on the other hand, generates the \emph{normal form}:\ a (weakly) smaller matching market with the same set of stable matchings that is, in a sense, ``boxed in'' by the two extreme stable matchings.
IDUA only generates a stable matching in the event that the original market possesses exactly one in which case the normal form is the unique stable matching and nothing more (see the main result of \cite{GutinNeary:2023:GEB}).

\subsubsection{Deferred acceptance}

In a run of DA, one side of the market makes proposals that the other side reacts to.
We frame it such that men propose and women respond.

DA begins by initialising all men as single.
Each round of the algorithm has three steps.
First, each single man proposes to the woman currently top of his preference list.
Second, each woman that received at least one proposal tentatively accepts the top proposer and rejects the other proposers.
Third, each man who was rejected truncates his preference list by removing from it the woman that he proposed to and is deemed single for the next round.

\setcounter{algorithm}{0}
\begin{algorithm}[!ht]
  \caption{DA algorithm}\label{algo:DA}\,
  
  Initialise all men as single.
  Each round of the algorithm proceeds as follows:

  \medskip
  \noindent
  1. Each single man proposes to his most preferred woman who has not yet rejected him.\\
  2. Each woman with at least one proposal tentatively accepts her top proposer and rejects all \red{\textbf{proposers}} that she ranks below him.\\
  3. Any man who is not matched becomes single.

  \medskip
  \noindent
  When there is a round in which no man is rejected, return the current pairs.
\end{algorithm}

That a stable matching exists for every two-sided matching market is a consequence of the following result.

\begin{theorem*}[\cite{GaleShapley:1962:AMM}]\label{thm:GaleShapley}
  The output of the man-proposing deferred acceptance algorithm is a stable matching.
\end{theorem*}

\subsubsection{Iterated deletion of unattractive alternatives}

A matching is deemed unstable if there exists a blocking pair relative to it.
According to this view, a blocking pair generates instability because it improves the fortunes of two participants who are not currently matched.
But \cite{GutinNeary:2023:GEB} and \cite{GutinNeary:2024:arXiv} put forward an alternative interpretation of instability.
They argue that the cause of instability can be attributed to issues with pairs that are contained within a matching -- specifically the two pairs that contribute to the breakaway couple.
These pairs can be thought of as ``weak'' relative to the matching.

A key point is that some pairs are always weak no matter the rest of the matching.
To see how, suppose that woman $\wom{}$ is man $\man{}$'s dream partner.
Then there cannot exist a stable matching in which $\wom{}$ is paired with a man that she ranks below $\man{}$.
Why?
Because $\wom{}$ could always (at least) propose to breakaway with $\man{}$ and $\man{}$ would certainly accept.
If stability is a requirement, then the analyst can delete pairs of the form described above.
That is, men that $\wom{}$ ranks below $\man{}$ can be removed from $\wom{}$'s preference list and for each of these men $\wom{}$ can be removed from his preference list.

The above motivates the notion of \emph{unattractive alternatives}.

\begin{definition}[Unattractive alternatives]\label{def:unattractive}
  We say that man $\man{}$ is an {unattractive alternative} to woman $\wom{}$, if there is some man $\man{}' \neq \man{}$ such that (i) $\man{}' \pref{\wom{}} \man{}$, and (ii) $\wom{} \pref{\man{}'} \wom{}'$ for all $\wom{}' \neq \wom{}$.
  A mirror-image statement describes when a woman is an unattractive alternative to a man.
  Finally, we say that $\wom{}$ is an {unattractive alternative} to man $\man{}$ whenever $\man{}$ is an {unattractive alternative} to woman $\wom{}$ (and vice versa).
\end{definition}

Deleting unattractive alternatives leaves behind a smaller matching market that, by definition, has the same set of stable matchings as the original.
More importantly, the deletion operation can iterate.
To see how, suppose that woman $\wom{}$ is the top choice for two men, $\man{1}$ and $\man{2}$, and suppose further that $\man{1} \pref{\wom{}} \man{2}$.
While $\wom{}$'s preference list will be truncated from $\man{1}$ on down, note that $\man{2}$'s preference list will be truncated from the top since $\wom{}$ is removed.
This means that $\man{2}$ now has a new most-preferred woman, say $\wom{}'$, since a match with $\wom{}$ ain't happening.
This may provide a new outside option for $\wom{}'$ which might in turn allow $\wom{}'$ to truncate her preference lists in a way that she was unable to before she became $\man{2}$'s top choice.

The above describes the beginnings of the \emph{iterated deletion of unattractive alternatives} (IDUA).
This is a procedure that repeatedly prunes redundant information from the preference lists as described above.
Technically it continually deletes unattractive alternatives from preference lists until there remains no market participant who views any other as unattractive.
When the procedure stops, the (sub)matching market that remains is referred to as the \emph{normal form}.
That is, the normal form $P^{*}$ is what remains when no further deletions are possible.\footnote{Section 2.3 of \cite{GutinNeary:2023:GEB} argues that IDUA is the matching market analog of the \emph{iterated deletion of dominated strategies} procedure for strategic games. The reason for the parallel is that both leave behind a smaller mathematical object without changing the set of ``solutions''.}

\begin{definition}[The iterated deletion of unattractive alternatives (IDUA)]\label{def:IDUAe}
  Given an instance of the matching problem $P = \big(\set{\pref{\man{}}}_{\man{}\in \Men}, \set{\pref{\wom{}}}_{\wom{}\in \Wom}\big)$, we define $\pref{\man{}}^{0} \, := \, \pref{\man{}}$ and $\pref{\wom{}}^{0} \, := \, \pref{\wom{}}$, and for each $k \geq 1$, form the matching (sub)problem $P^{k} = \big(\{\pref{\man{}}^{k}\}_{\man{}\in \Men}, \{ \pref{\wom{}}^{k}\}_{\wom{}\in \Wom } \big)$ where for every man $\man{}$ and every woman $\wom{}$,
  \begin{equation}\label{eq:iterative}
    \begin{aligned}
      \pref{\man{}}^{k} & = \set{\wom{}  \, \big| \,\, \wom{} \in \set{\pref{\man{}}^{k-1}} \, \text{ and } \, \man{} \in \set{\pref{\wom{}}^{k-1}} }, \text{ and } \\
      \pref{\wom{}}^{k} & = \set{\man{} \,  \big| \,\, \man{} \in \set{\pref{\wom{}}^{k-1}} \, \text{ and } \, \wom{} \in \set{\pref{\man{}}^{k-1}}}.
    \end{aligned}
  \end{equation}
  Finally, define the {\em normal form} of matching problem $P$, $P^{*}$, as $P^{k^{*}}$ where $k^*$ is the minimum $k$ such that $P^{k+1} = P^{k}$.
  Men's preferences on the normal form are denoted $\set{\pref{\man{}}^{*}}$ and those of women by $\{\pref{\wom{}}^{*}\}$.
\end{definition}

The following lemma confirms that, for an analyst interest in the set of stable matchings, restricting attention to the normal form is sufficient.

\begin{lemma}[\cite{BalinskiRatier:1997:,GutinNeary:2023:GEB}] \label{lemma:iduaSameMatchingsE}
  The iterated deletion of unattractive alternatives does not change the set of stable matchings.  That is, $P$ and its associated normal form, $P^{*}$, contain exactly the same set of stable matchings.
\end{lemma}

The next lemma shows that the normal form is ``boxed in'' by the man-optimal and woman-optimal stable matchings.
Before stating the lemma, we introduce some notation.
Given a matching problem's normal form $P^*$, for each man $\man{}$, let $\tau(\man{})$ denote the woman at the top of $\man{}$'s truncated preference list in $P^*$, and similarly, for each $\wom{}$, let $\tau(\wom{})$ denote the man at the top of $\wom{}$'s truncated list.

\begin{lemma}[\cite{BalinskiRatier:1997:,GutinNeary:2023:GEB}]\label{lemma:extremal}
  Let $P$ be an instance of the matching problem and let $P^*$ denote its normal form.
  The following two collections of pairs, $\match_{\Men}$ and $\match_{\Wom}$, are stable matchings in $P$.
  \begin{center}
    $\begin{array}{rcl} \vspace{0.2cm}
        \match_{\Men} & = & \Set{\big(\man{1}, \tau(\man{1})\big), \dots, \big(\man{n}, \tau(\man{n})\big)} \\
        \match_{\Wom} & = & \Set{\big(\tau(\wom{1}), \wom{1}\big), \dots, \big(\tau(\wom{n}), \wom{n}\big)} \\
      \end{array}$
  \end{center}
\end{lemma}

Consider the collection of pairs $\match_{\Men}$.
To see that $\match_{\Men}$ is a matching, observe that every man must have a different most preferred woman in the normal form.
If not, then two men have the same most-preferred woman.
But this would mean that the IDUA procedure is not yet finished and so the normal form has not yet been reached.
To see that the matching $\match_{\Men}$ is stable, note that every man is paired with his most preferred feasible woman (any preferred woman is infeasible because she is no longer on his preference list).
Such a matching must be stable because no man is willing to swap and so there can be no blocking pairs.

The matchings $\match_{\Men}$ and $\match_{\Wom}$ are typically referred to as the \emph{man-optimal stable matching} and the \emph{woman-optimal stable matching} respectively.


\section{Accelerated deferred acceptance}\label{sec:algorithm}

In this section we introduce the \emph{accelerated deferred acceptance} algorithm.
ADA is strikingly similar to the DA algorithm of \citeauthor{GaleShapley:1962:AMM}.
The key difference is that ADA moves at greater pace through the market because preferences are truncated in response to proposals much as they are in IDUA.
In fact, accelerated deferred acceptance can be interpreted as a one-sided variant of IDUA.

Section~\ref{ssec:ADA} defines accelerated deferred acceptance and confirms that it generates the same outcome as DA and that it is strategy proof for the proposing side.
In Section~\ref{ssec:example} we show that accelerated deferred acceptance has practical improvements over DA:\ it requires fewer proposals (Theorem~\ref{thm:ADAproposals}), it terminates in fewer rounds (Theorem~\ref{thm:ADArounds}), and all tentative matches take place no later (Theorem~\ref{thm:ADAearlier}).
These features are illustrated by Example~\ref{ex:bothAlgos}, where both algorithms are run on the same market.

\subsection{The algorithm}\label{ssec:ADA}

The accelerated deferred acceptance algorithm is similar to DA except that whenever a woman receives a new a proposal, she rejects all men who she ranks below the most preferred proposing man, not just those who have also proposed to her.\footnote{We remind the reader that the key difference between \citeauthor{GaleShapley:1962:AMM}'s DA algorithm (Algorithm~\ref{algo:DA}) and our accelerated deferred acceptance (Algorithm~\ref{algo:ADA}) is one word:\ accelerated deferred acceptance replaces the word ``proposers'' in DA with the word ``men''. For emphasis we have typeset the updated word in \textbf{\red{red}}.}
Some of these rejections may be pre-emptive in that a subset of men are forbidden from ever proposing to certain women.
Some of these men might have ultimately proposed while others might have never got around to it; but all are prevented from doing so.

\begin{algorithm}
  \caption{Accelerated deferred acceptance algorithm}\label{algo:ADA}\,
  
  Initialise all men as single.
  Every round of the algorithm proceeds as follows:

  \medskip
  \noindent
  1. Each single man proposes to his most preferred woman who has not yet rejected him.\\
  2. Each woman with at least one proposal tentatively accepts her top proposer and rejects all \textbf{\red{men}} that she ranks below him.\\
  3. Any man who is not matched becomes single.

  \medskip
  \noindent
  When there is a round in which no man is rejected, return the current pairs.
\end{algorithm}

Our main result in this section, Theorem~\ref{thm:ADAstable}, confirms that accelerated deferred acceptance finds the same stable matching as that found by DA.
To prove the result, first we show that accelerated deferred acceptance always terminates, that it terminates at a matching, and that the matching arrived at is stable.
We then show that the man-proposing accelerated deferred acceptance finds the man-optimal stable matching, thereby confirming that the output is equivalent to that of DA.

The following two observations are used in proving the result.

\begin{observation}\label{obs:decrease}
  Men propose to women in decreasing order of preference.
\end{observation}
\begin{observation}\label{obs:never}
  Once a woman is matched, she never becomes unmatched.
\end{observation}

Observation~\ref{obs:decrease} is a restatement of the fact that each man works down his preference list under ADA.
However, a man may not work incrementally down his list as with DA, because it may be that the woman who would be top of his ``as yet unproposed to'' list has already pre-emptively rejected him.
By Observation~\ref{obs:never}, women only ever ``trade up''.
Again, while the same statement holds for DA, there is a difference.
For ADA, whenever a woman receives a new proposal in a given round she is guaranteed of trading up because a proposal can only arrive from a man who she prefers to her current match.

We now state the result.
All proofs can be found in Appendix~\ref{app:proofs}.

\begin{theorem}\label{thm:ADAstable}
The man-proposing accelerated deferred acceptance returns the same matching as the man-proposing deferred acceptance algorithm.
\end{theorem}

In the context of two-sided matching markets, an \emph{allocation rule} is a mapping from the reported preferences of all $2n$ market participants to the set of matchings.
An allocation rule is said to be \emph{strategy proof} if it is a weakly dominant strategy for participants to report their preferences truthfully.
A natural concern with any allocation rule is whether or not it is {strategy proof}.
That is, in the event that preferences are private information participants not known to a market designer, is the truthful reporting of preferences incentive compatible with a direct mechanism that generates the stable matching found by accelerated deferred acceptance?
A mechanism that is not strategy proof is game-able, and many things can go wrong with game-able systems.

The DA algorithm is strategy proof for the proposing side \citep{DubinsFreedman:1981:AMM,Roth:1982:MOR}.\footnote{For a concise proof see \citet[\S 10.4, page 258]{NisanRoughgarden:2007:}.}
Given that accelerated deferred acceptance always generates the same output as DA, it holds that misreporting one's true preferences to a mechanism based on men-proposing can never lead to a better outcome for a man.
That is, we get the following result for free.

\begin{theorem}\label{thm:ADAstrategyProof}
The allocation rule associated with the men-proposing accelerated deferred acceptance algorithm is strategy proof for the men.
\end{theorem}

Theorem~\ref{thm:ADAstrategyProof} is a statement about the proposers, in this case the men.
The mechanism associated with men-proposing accelerated deferred acceptance is not strategy proof for the women.
The proof is via a counterexample.
Consider a two-sided market with two or more stable matchings so that the extreme stable matchings $\match_\Men$ and $\match_\Wom$ are different.
For the women, the matching $\match_\Wom = \set{\big(\tau(\wom{1}), \wom{1}\big), \dots, \big(\tau(\wom{n}), \wom{n}\big)}$ Pareto dominates $\match_\Men$.
Therefore, if every woman $\wom{i}$ submits a truncated preference list to the mechanism containing only the element $\tau(\wom{i})$, then the mechanism will return $\match_\Wom$ which for every woman is (weakly) preferable to $\match_\Men$.

\subsection{Fewer proposals, fewer rounds, and earlier pairings}\label{ssec:example}

In this section we compare and contrast both algorithms along various criteria.
We begin with an example to illustrate how ADA typically ``moves more quickly'' through a matching market than DA.

\begin{example}\label{ex:bothAlgos}
  Consider a one-to-one two-sided matching problem with five men, $\Men = \set{\man{1}, \man{2}, \man{3}, \man{4}, \man{5}}$, and five women, $\Wom = \set{\wom{1}, \wom{2}, \wom{3}, \wom{4}, \wom{5}}$.
  The preference lists for each man and each woman are presented below, with the participants on the other side of the market listed in order of decreasing preference.
  \begin{align}
     & \wom{1}: \, \man{5} \, , \man{4} \, , \man{1}, \,  \man{2}, \, \man{3}  \hspace{.1in}  & \blc{\blue{\man{1}}}: & \, \wom{1} \, , \wom{2} \, , \wom{3} \, , \wom{4} \, , \wom{5} 	\nonumber            \\
     & \wom{2}: \,  \man{1} \, , \man{3} \, , \man{2}, \, \man{4}, \, \man{5}  \hspace{.1in}  & \blc{\blue{\man{2}}}: & \,\, \wom{1} \, , \wom{4} \, , \wom{5} \, , \wom{2}, \, \wom{3} \nonumber            \\
     & \wom{3}: \,  \man{5} \, , \man{4} \, , \man{3}, \,  \man{2}, \, \man{1}  \hspace{.1in} & \blc{\blue{\man{3}}}: & \,\, \wom{1} \, , \wom{4} \, , \wom{3} \, , \wom{5}, \, \wom{2}	\label{eq:prefLists} \\
     & \wom{4}: \,  \man{4} \, , \man{2} \, , \man{1}, \,  \man{3}, \, \man{5}  \hspace{.1in} & \blc{\blue{\man{4}}}: & \,\, \wom{4} \, , \wom{2} \, , \wom{3} \, , \wom{1}, \, \wom{5}	\nonumber            \\
     & \wom{5}: \,  \man{5} \, , \man{1} \, , \man{3}, \,  \man{4}, \, \man{2}  \hspace{.1in} & \blc{\blue{\man{5}}}: & \,\, \wom{5} \, , \wom{4} \, , \wom{1} \, , \wom{2}, \, \wom{3} \nonumber
  \end{align}

  We now show how both DA and accelerated deferred acceptance operate on the market above.
  In the interest of space we will only present how both algorithms operate in the first two rounds.
  In fact, two rounds is enough for accelerated deferred acceptance to terminate but not so for DA as it requires four rounds. 
  How DA continues to completion is spelled out in Appendix~\ref{app:example}.

Both DA and accelerated deferred acceptance begin with initialising all men as single.
To indicate this in \eqref{eq:prefLists}, we have written men in \blue{blue} if they are active in ADA and circled a man in a blue circle if he is active in DA.
  All men are active in Round 1 so each man $\man{i}$ is written as $\blc{\blue{\man{i}}}$.

  In the first round each man approaches his top ranked woman.
  This is depicted in the Table~\ref{table:R1} below.
  Below each woman are two columns that convey whether the woman is proposed to under DA or ADA.
  The column in \gray{gray} is for DA and the column in \green{green} is for ADA.
  When woman $\wom{}$ receives multiple proposals we order the suitors in the column from top to bottom in order of decreasing preference for $\wom{}$.
  In this example, woman $\wom{1}$ received multiple proposers in Round 1.
  She will tentatively accept the top proposer, man $\man{1}$.
  All tentative acceptances are represented by labelling the accepted man with an asterix.
  \noindent
  \extrarowheight=\aboverulesep
  \addtolength{\extrarowheight}{\belowrulesep}
  \aboverulesep=0.1pt
  \belowrulesep=0.1pt
\begin{table}[h!]
\caption{proposals and tentative acceptances in Round 1.}
  \begin{tabular}{L{0.5cm}C{0.5cm}C{0.5cm}||C{0.5cm}C{0.5cm}||C{0.5cm}C{0.5cm}||C{0.5cm}C{0.5cm}||C{0.5cm}C{0.5cm}}
     & \multicolumn{2}{C{1.4cm}||}{$\wom{1}$} & \multicolumn{2}{C{1.4cm}||}{$\wom{2}$} & \multicolumn{2}{C{1.4cm}||}{$\wom{3}$} & \multicolumn{2}{C{1.4cm}||}{$\wom{4}$} & \multicolumn{2}{C{1.4cm}}{$\wom{5}$}                                                                                                                                                                     \\
    \cmidrule{2-11} \morecmidrules \cmidrule{2-11}
     & \cellcolor{lightgray}$\man{1}^*$       & \cellcolor{greengray}$\man{1}^*$       & \cellcolor{lightgray}                  & \cellcolor{greengray}                  & \cellcolor{lightgray}                & \cellcolor{greengray} & \cellcolor{lightgray}$\man{4}^*$ & \cellcolor{greengray}$\man{4}^*$ & \cellcolor{lightgray}$\man{5}^*$ & \cellcolor{greengray}$\man{5}^*$ \\

     & \cellcolor{lightgray}$\man{2}$         & \cellcolor{greengray}$\man{2}$         & \cellcolor{lightgray}                  & \cellcolor{greengray}                  & \cellcolor{lightgray}                & \cellcolor{greengray} & \cellcolor{lightgray}            & \cellcolor{greengray}            & \cellcolor{lightgray}            & \cellcolor{greengray}            \\

     & \cellcolor{lightgray}$\man{3}$         & \cellcolor{greengray}$\man{3}$         & \cellcolor{lightgray}                  & \cellcolor{greengray}                  & \cellcolor{lightgray}                & \cellcolor{greengray} & \cellcolor{lightgray}            & \cellcolor{greengray}            & \cellcolor{lightgray}            & \cellcolor{greengray}            \\
  \end{tabular}
  \label{table:R1}
\end{table}
  \bigskip

  Table~\ref{table:R1} listed proposals and acceptances in Round 1.
  Now let us turn to rejections.
  This is the first point at which the two algorithms diverge.
Some reflection reveals that ADA always starts out with weakly more rejections than DA. 
This is the case because any man rejected under DA is rejected under ADA, but pre-emptive rejections may also be handed out.

We represent rejections by updating the preference lists as in \eqref{eq:prefLists1}.
We denote rejections that occur under ADA in \red{red}.
Similarly, a rejection that occurs under DA is depicted by putting a red box around the relevant party.
For example, since man $\man{2}$ was rejected by woman $\wom{1}$ in both algorithms, we write $\rds{\red{\man{2}}}$ in $\wom{1}$'s preference list and write $\rds{\red{\wom{1}}}$ in $\man{2}$'s preference list.

Note the preference lists of $\wom{4}$ and $\wom{5}$ in \eqref{eq:prefLists1} .
Each woman was proposed to by her top ranked male, $\man{4}$ for $\wom{4}$ and $\man{5}$ for $\wom{5}$.
Both algorithms therefore include the two tentative pairs $(\man{4},\wom{4})$ and $(\man{5},\wom{5})$.
But for ADA there are pre-emptive rejections handed out by $\wom{4}$ and $\wom{5}$.
It is for this reason that some entries on their respective preference lists are in \red{red}.
  \begin{align}
     & \wom{1}: \, \man{5} \, , \man{4} \, , \man{1}, \,  \rds{\red{\man{2}}}, \, \rds{\red{\man{3}}}  \hspace{.1in}  & \man{1}:              & \, \wom{1} \, , \wom{2} \, , \wom{3} \, , \red{\wom{4}} \, , \red{\wom{5}}       \nonumber       \\
     & \wom{2}: \,  \man{1} \, , \man{3} \, , \man{2}, \, \man{4}, \, \man{5}  \hspace{.1in}                          & \blc{\blue{\man{2}}}: & \,\, \rds{\red{\wom{1}}} \, , \red{\wom{4}}\, , \red{\wom{5}} \, , \wom{2}, \, \wom{3} \nonumber  \\
     & \wom{3}: \,  \man{5} \, , \man{4} \, , \man{3}, \,  \man{2}, \, \man{1}  \hspace{.1in}                         & \blc{\blue{\man{3}}}: & \,\, \rds{\red{\wom{1}}} \, , \red{\wom{4}} \, , \wom{3} \, , \red{\wom{5}}, \, \wom{2} \label{eq:prefLists1} \\
     & \wom{4}: \,  \man{4} \, , \red{\man{2}} \, , \red{\man{1}}, \,  \red{\man{3}}, \, \red{\man{5}}  \hspace{.1in} & \man{4}:              & \,\, \wom{4} \, , \wom{2} \, , \wom{3} \, , \wom{1}, \, \red{\wom{5}}      \nonumber             \\
     & \wom{5}: \,  \man{5} \, , \red{\man{1}} \, , \red{\man{3}}, \,  \red{\man{4}}, \, \red{\man{2}}  \hspace{.1in} & \man{5}:              & \,\, \wom{5} \, , \red{\wom{4}} \, , \wom{1} \, , \wom{2}, \, \wom{3}          \nonumber         
  \end{align}

Let us now move to Round 2.
The only single men at the beginning of this round are $\man{2}$ and $\man{3}$.
DA stipulates that both men propose to the top ranked as-yet-unproposed-to woman on their list.
From \eqref{eq:prefLists1}, we can see that for both men this is $\wom{4}$.
For ADA there is a difference.
Consider $\man{2}$.
During Round 1, he received a pre-emptive rejection from $\wom{4}$ and $\wom{5}$, conveyed by writing $\red{\wom{4}}$ and $\red{\wom{5}}$ in the preference list of $\man{2}$ in \eqref{eq:prefLists1}.
Therefore $\man{2}$ proposes to $\wom{2}$ in Round 2.

The full list of proposals in Round 2 are as in Table~\ref{table:R2} below.

\begin{table}[h!]
\caption{proposals and tentative acceptances in Round 2.}
  \begin{tabular}{L{0.5cm}C{0.5cm}C{0.5cm}||C{0.5cm}C{0.5cm}||C{0.5cm}C{0.5cm}||C{0.5cm}C{0.5cm}||C{0.5cm}C{0.5cm}}
     & \multicolumn{2}{C{1.4cm}||}{$\wom{1}$} & \multicolumn{2}{C{1.4cm}||}{$\wom{2}$} & \multicolumn{2}{C{1.4cm}||}{$\wom{3}$} & \multicolumn{2}{C{1.4cm}||}{$\wom{4}$} & \multicolumn{2}{C{1.4cm}}{$\wom{5}$}                                                                                                                                                                        \\
    \cmidrule{2-11} \morecmidrules \cmidrule{2-11}
     & \cellcolor{lightgray}$\man{1}^{*}$         & \cellcolor{greengray}$\man{1}^{*}$         & \cellcolor{lightgray}                  & \cellcolor{greengray}$\man{2}^{*}$         & \cellcolor{lightgray}                & \cellcolor{greengray}$\man{3}^{*}$ & \cellcolor{lightgray}$\man{4}^{*}$   & \cellcolor{greengray}$\man{4}^{*}$ & \cellcolor{lightgray}$\man{5}^{*}$ & \cellcolor{greengray}$\man{5}^{*}$ \\

     & \cellcolor{lightgray}                  & \cellcolor{greengray}                  & \cellcolor{lightgray}                  & \cellcolor{greengray}                  & \cellcolor{lightgray}                & \cellcolor{greengray}          & \cellcolor{lightgray}$\man{3}$ & \cellcolor{greengray}          & \cellcolor{lightgray}          & \cellcolor{greengray}          \\

     & \cellcolor{lightgray}                  & \cellcolor{greengray}                  & \cellcolor{lightgray}                  & \cellcolor{greengray}                  & \cellcolor{lightgray}                & \cellcolor{greengray}          & \cellcolor{lightgray}$\man{2}$ & \cellcolor{greengray}          & \cellcolor{lightgray}          & \cellcolor{greengray}          \\
  \end{tabular}
  \label{table:R2}
\end{table}
  \bigskip

Note from Table~\ref{table:R2} above that ADA is finished at the end of Round 2.
It has generated a collection of five tentative pairs $\set{(\man{i}, \wom{i})}_{i=1}^{5}$ that make up a matching. Theorem~\ref{thm:ADAstable} assures that this matching is stable.

In contrast to ADA, the DA algorithm does not terminate after two rounds.
In Round 2 of DA, both $\man{2}$ and $\man{3}$ proposed to woman $\wom{4}$.
But $\wom{4}$ was already tentatively paired with $\man{4}$ who she prefers over each.
So $\wom{4}$ rejects both proposals.
Both $\man{2}$ and $\man{3}$ will propose to somebody else in Round 3.
The preference lists are updated as in \eqref{eq:prefLists2}.
After 4 rounds, DA ultimately terminates at the same stable matching as that found by ADA.
The details are provided in Appendix~\ref{app:example}.

  \begin{align}
     & \wom{1}: \, \man{5} \, , \man{4} \, , \man{1}, \,  \rds{\red{\man{2}}}, \, \rds{\red{\man{3}}}  \hspace{.1in}              & \man{1}:              & \, \wom{1} \, , \wom{2} \, , \red{\wom{3}} \, , \red{\wom{4}} \, , \red{\wom{5}}        \nonumber           \\
     & \wom{2}: \,  \man{1} \, , \man{3} \, , \man{2}, \, \red{\man{4}}, \, \red{\man{5}}  \hspace{.1in}                          & \blc{{\man{2}}}: & \,\, \rds{\red{\wom{1}}} \, , \rds{\red{\wom{4}}}\, , \red{\wom{5}} \, , \wom{2}, \, \red{\wom{3}} \nonumber \\
     & \wom{3}: \,  \man{5} \, , \man{4} \, , \man{3}, \,  \red{\man{2}}, \, \red{\man{1}}  \hspace{.1in}                         & \blc{{\man{3}}}: & \,\, \rds{\red{\wom{1}}} \, , \rds{\red{\wom{4}}} \, , \wom{3} \, , \red{\wom{5}}, \, \wom{2} \label{eq:prefLists2}     \\
     & \wom{4}: \,  \man{4} \, , \rds{\red{\man{2}}} \, , \red{\man{1}}, \,  \rds{\red{\man{3}}}, \, \red{\man{5}}  \hspace{.1in} & \man{4}:              & \,\, \wom{4} \, , \red{\wom{2}} \, , \wom{3} \, , \wom{1}, \, \red{\wom{5}}               \nonumber         \\
     & \wom{5}: \,  \man{5} \, , \red{\man{1}} \, , \red{\man{3}}, \,  \red{\man{4}}, \, \red{\man{2}}  \hspace{.1in}             & \man{5}:              & \,\, \wom{5} \, , \red{\wom{4}} \, , \wom{1} \, , \red{\wom{2}}, \, \wom{3}         \nonumber               
  \end{align}

\end{example}

Let us now make some observations about the example.

After two rounds the number of proposals made by DA and ADA must be the same since each generates $n$ proposals in the first round, and all males rejected in the first round make a new proposal in the second.
This holds for our example with both $\man{2}$ and $\man{3}$ single at the beginning of Round 2.

Note that in Round 2 of the DA algorithm, all proposals made were rejected. 
We refer to a round of this kind as an \emph{idle round}, since the collection of matched pairs did not update.
The ADA algorithm cannot have an idle round.
In Section~\ref{sec:simulations} we will show that DA can have many.

The structure of the preference lists upon termination of both algorithms is worth noting.
DA is contiguous from the top down for men and potentially scattered for women. ADA is the opposite: contiguous from the bottom up for women and potentially scattered for men.

Let us now consider our measures of efficiency. 
We begin with the number of proposals needed for each algorithm to terminate.
From the analysis above we see that ADA required seven proposals to identify the man-optimal stable matching: 5 proposals in Round 1 and two proposals in Round 2.
From Appendix~\ref{app:example}, we see that DA required ten proposals.
It turns out that ADA requiring fewer proposals than DA is not a function of the specifics of Example~\ref{ex:bothAlgos}.
Rather, this is a general property that always holds, as confirmed by the following result that is proved in Appendix~\ref{app:proofs}.

\begin{theorem}\label{thm:ADAproposals}
Accelerated deferred acceptance never requires more proposals than DA.
\end{theorem}

Our next efficiency measure was number of rounds required to completion.
In Example~\ref{ex:bothAlgos}, ADA needed two rounds whereas DA required four rounds, one of which was an idle round.
Again, this efficiency gain is not an artefact of Example~\ref{ex:bothAlgos}.
Rather, we have the following assurance.

\begin{theorem}\label{thm:ADArounds}
  Accelerated deferred acceptance always terminates in weakly fewer rounds than DA.
\end{theorem}

In fact, Theorem~\ref{thm:ADArounds} above follows almost immediately from the following, stronger result.

\begin{theorem}\label{thm:ADAearlier}
Each proposal that is made when running ADA takes place in a weakly earlier round than for DA.
\end{theorem}

It is clear that ADA is weakly ``ahead'' of DA after only one round of each algorithm since both contain the same number of direct rejections and ADA may contain some pre-emptive rejections too.
Theorem~~\ref{thm:ADAearlier} confirms that ADA stays.
The proof of Theorem~\ref{thm:ADAearlier} is in Appendix~\ref{app:proofs}.

Theorem~\ref{thm:ADArounds} follows from Theorem~\ref{thm:ADAearlier} because all final pairs match no later for ADA.
That is, ADA terminates when the last final pair is matched and this happens as the result of a proposal.


\section{Computational experiments}\label{sec:simulations}

Our theoretical results in Section~\ref{sec:algorithm} show that ADA is more efficient than DA in certain ways. 
However, the results are completely silent on the extent of the efficiency gains.
It is reasonable to suppose that these gains might vary with the structure of preferences and market size.
The computational experiments described in this section are designed to address this.

Specifically, the objectives of our computational study are as follows:
\begin{itemize}
	\item
	Study the number of rounds and the number of proposals generated by ADA compared to DA.
	
	\item
	Study the proportion of final pairs by round, which is a new metric that captures when each final pair first matched.
	
	\item
	Compare the running times of the two algorithms when the preference lists are pre-defined.
\end{itemize}



Section~\ref{ssec:implementation} defines the pseudo-random instance generator and Sections~\ref{ssec:rounds}, \ref{ssec:proposals}, \ref{ssec:final_pairs} and \ref{ssec:running_time} present our findings.

\subsection{Instance generator}\label{ssec:implementation}

Since the number of instances of a two-sided matching problem of size $n$ is $(n!)^{2n}$, an exhaustive exploration of how the algorithms perform on every instance is intractable from $n=5$ and up.
For this reason, we propose a novel instance generator that allows one to sample random instances for any market size $n$ where the bias in the sampling is controlled by a parameter $c \in [0,1]$.

Let an instance be a tuple $(\boldsymbol{W}, \boldsymbol{M})$, where $\boldsymbol{W} = (w_{i,j})$ and $\boldsymbol{M} = (m_{i,j})$ are $n \times n$ matrices.
Each row of each of these matrices is a permutation of $(1, 2, \ldots, n)$.

\begin{definition}[Instance generator]
  Our instance generator takes two parameters:\ a non-negative integer $n$ for the size of the instance and a coefficient $c \in[0,1]$ that controls the bias in the sampling distribution over preferences of participants on the same side of the market.\footnote{We assume that the bias is the same for both sides. This can be relaxed.}
  The generator works as follows.
  \begin{enumerate}
    \item
          Create a random permutation $\boldsymbol{p}$ of values $(0, 1, \ldots, n-1)$.

    \item
          For each $i \in \{ 1, 2, \ldots, n \}$, create a random vector $\boldsymbol{v}_{i} \in \mathbb{R}^n$, where each element of $\boldsymbol{v}_{i}$ is in the range $[0, n-1]$.

    \item
          Calculate $\boldsymbol{u}_{i} = (1 - c) \boldsymbol{v}_i + c \boldsymbol{p}$.
	Let $(m_{i,1}, m_{i,2}, \ldots, m_{i,n})$ be the permutation that orders $\boldsymbol{u}_{i}$ in ascending order.

    \item
          Follow steps 1--3 to produce matrix $\boldsymbol{W}$.
  \end{enumerate}

\end{definition}

Observe that the matrices $\boldsymbol{M}$ and $\boldsymbol{W}$ are uniformly random when $c=0$.
This follows because each row is drawn uniformly from the set of all possible preferences and all the rows are independent.
Conversely, when $c = 1$, the preferences of all the men are identical and the preferences of all the women are identical (\cite{HolzmanSamet:2014:GEB} say that such preferences possess a ``universal ranking'').

\subsection{Number of rounds}
\label{ssec:rounds}


The number of rounds and the number of proposals are arguably the most important metrics as it is easy to imagine a market where there are costs associated with each round and each proposal.

\begin{figure}[htb]
  \centering
  \includegraphics[width=0.85\textwidth,trim=0.5cm 0.6cm 0.5cm 1cm,clip]{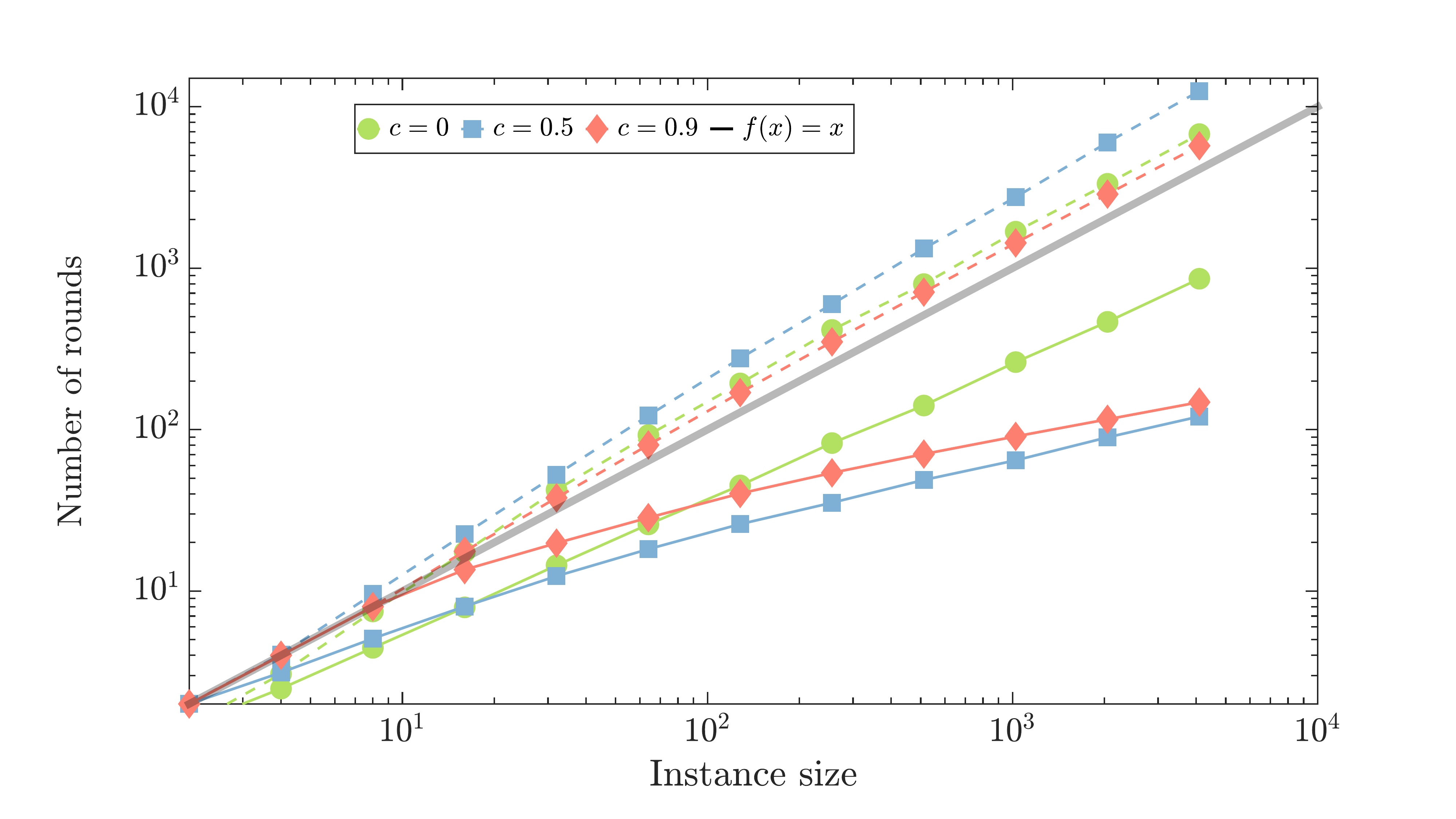}
  \caption{Number of rounds by $n$.  {DA: dashed lines; ADA: solid lines.}}
  \label{fig:rounds_by_n}
\end{figure}

Figure~\ref{fig:rounds_by_n} shows how the number of rounds changes with the the size of the market for three different values of the bias parameter: \emph{uniformly random} ($c=0$), \emph{moderately similar} ($c=0.5$), or \emph{similar} ($c=0.9$).
The values of $n$ change as $n = 2, 4, 8, \ldots, 4\,096$.
The lines are colour-coded according to value of $c$.
The dashed lines show the average rounds taken under DA, while the solid lines show corresponding results for ADA\@.

Note that, unless specified otherwise, each point in each figure in this section corresponds to a mean value over 1\,000 instances generated with identical instance generator parameter values but distinct pseudo-random number generator seeds.

For all three values of $c$, the total number of rounds for DA increases super-linearly with instance size. 
In contrast, the number of rounds always evolves sublinearly for ADA\@.
This difference leads to a substantial reduction in rounds for ADA, particularly for large instance sizes.

\begin{figure}[!ht]
  \centering
  \includegraphics[width=0.94\textwidth]{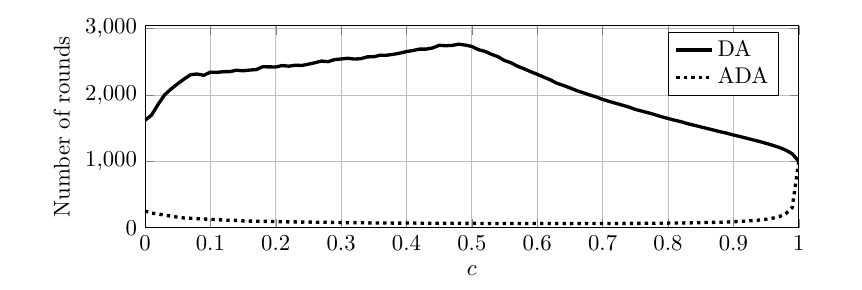}
%
%
%
%
\caption{Number of rounds as a function of $c$.}
\label{fig:rounds_by_c}
\end{figure}

It is evident from Figure~\ref{fig:rounds_by_n} that the number of rounds significantly depends on the value of $c$.
To investigate this dependency further, we run another set of experiments where we fixed $n = 1\,000$ and varied $c$ as $c = 0, 0.01, 0.02, \ldots, 1$.
The results are reported in Figure~\ref{fig:rounds_by_c} (for this experiment, we increased the number of instances per point to $10\,000$ as the high variance was causing significant noise).
As in Figure~\ref{fig:rounds_by_n}, the difference between the two algorithms is stark.
For example, when $c = 0$, ADA reduces the number of rounds by a factor of 7.
When $c = 0.5$, the scaling factor increases to 50.
In fact, the values of $c$ close to $c = 0.5$ seem to maximise the number of rounds for DA and minimise the number of rounds of ADA.

\begin{figure}[!ht]
  \centering
  \includegraphics[width=1.0\textwidth]{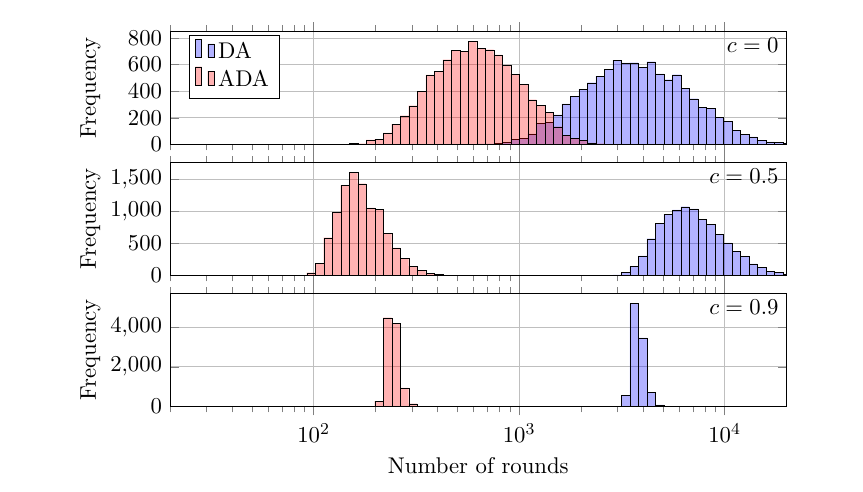}
\caption{The distributions of the number of rounds for $c = 0.0$, $c = 0.5$ and $c = 0.9$.}
\label{fig:rounds_distributions}
\end{figure}

As mentioned above, we found that the number of rounds by DA has a significant variation.
Figure~\ref{fig:rounds_distributions} shows the distributions of the number of rounds for DA and ADA for uniformly random, moderately similar and similar instances.
(Once again, we used 10\,000 instances for each value of $c$ for this experiment.)

%
%
%
%
%

An interesting observation from Figure~\ref{fig:rounds_distributions} is that the two distributions for DA and ADA are similar except shifted.
Since the scale of the horizontal axis is logarithmic, this means that the \emph{relative} variances for the two algorithms are similar, however the \emph{absolute} variances are very different.
In other words, the number of rounds of ADA is much more predictable than the number of rounds of DA, which could be an important strength of ADA.

\begin{figure}[!ht]
  \centering
  \includegraphics[width=1.0\textwidth]{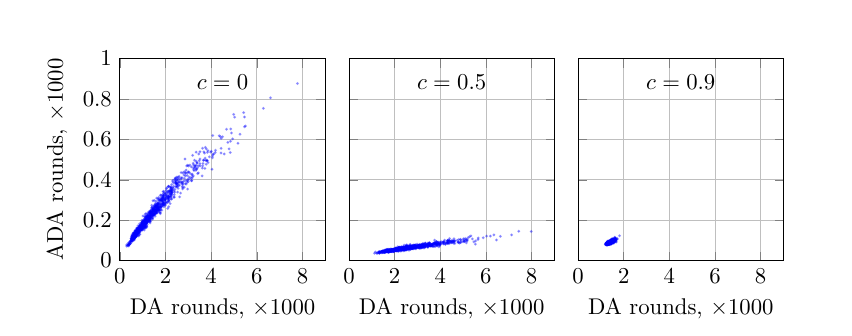}
\caption{The relation between the number of rounds for DA and ADA.}
\label{fig:relation}
\end{figure}

All the experiments so far demonstrated that ADA is superior on average to DA with respect to various statistics.
However, it is natural to question whether there exist instances that are actually particularly hard for ADA but relatively easy for DA.
To investigate this, we produced Figure~\ref{fig:relation} that shows how the number of rounds of DA is related to the number of rounds of ADA for each instance ($1\,000$ instances are used for each $c = 0, 0.5, 0.9$).

As evident from Figure~\ref{fig:relation}, the numbers of rounds of the two algorithms are highly correlated.
In other words, the instances that are hard for ADA are also hard for DA and vice versa.
Of course we cannot rule out that there exist instances for which DA performs similarly to ADA.
However, we have not observed any in our experiments.

\subsection{Number of proposals}
\label{ssec:proposals}

\begin{figure}[htbp]
  \centering
  \includegraphics[width=0.85\textwidth,trim=0.5cm 0.6cm 0.5cm 1cm,clip]{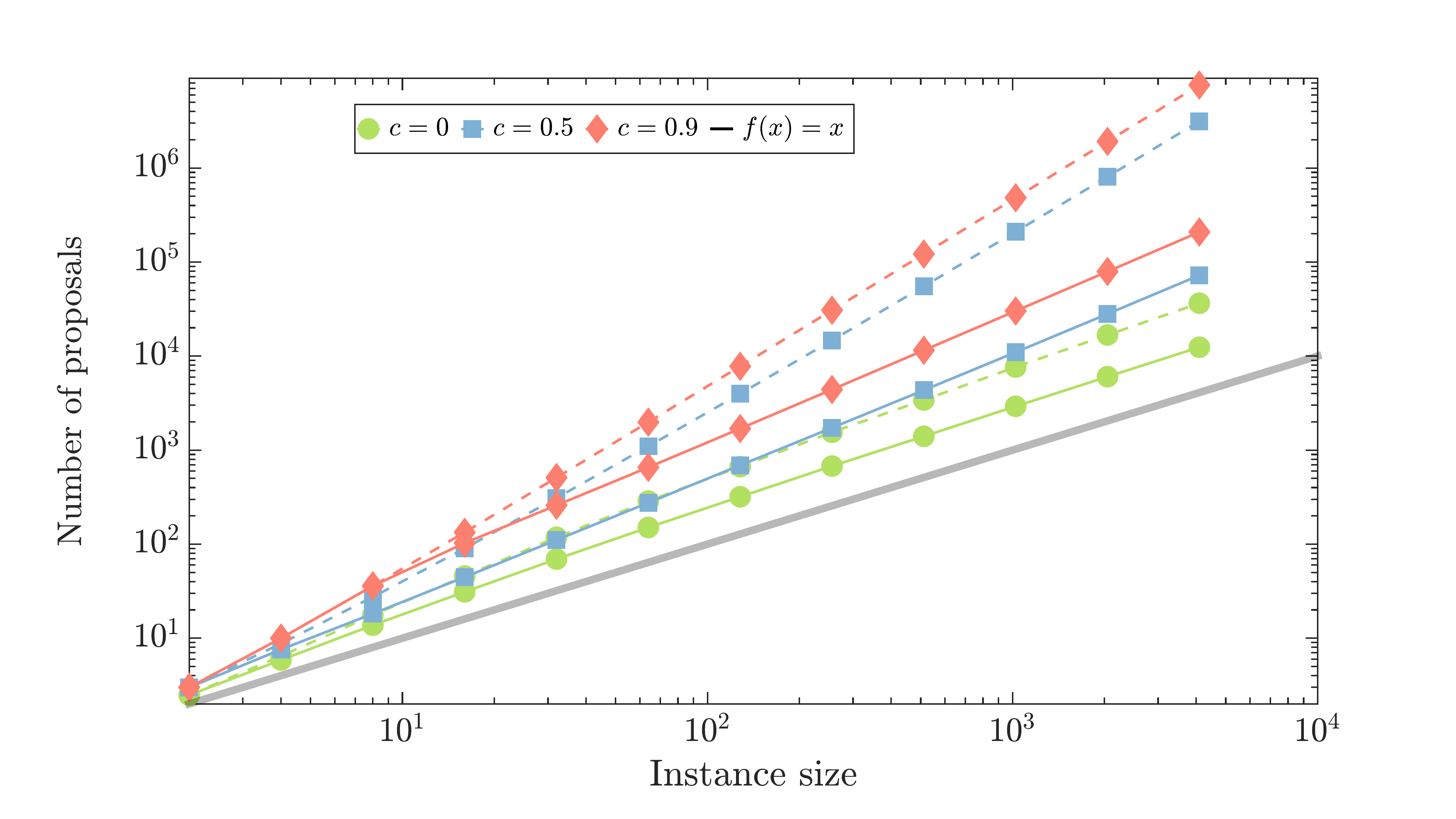}
  \caption{Number of proposals by $n$.}
  \label{fig:props_by_n}
\end{figure}

\begin{figure}[htbp]
  \centering
  \includegraphics[width=0.85\textwidth,trim=0.5cm 0.6cm 0.5cm 0.2cm,clip]{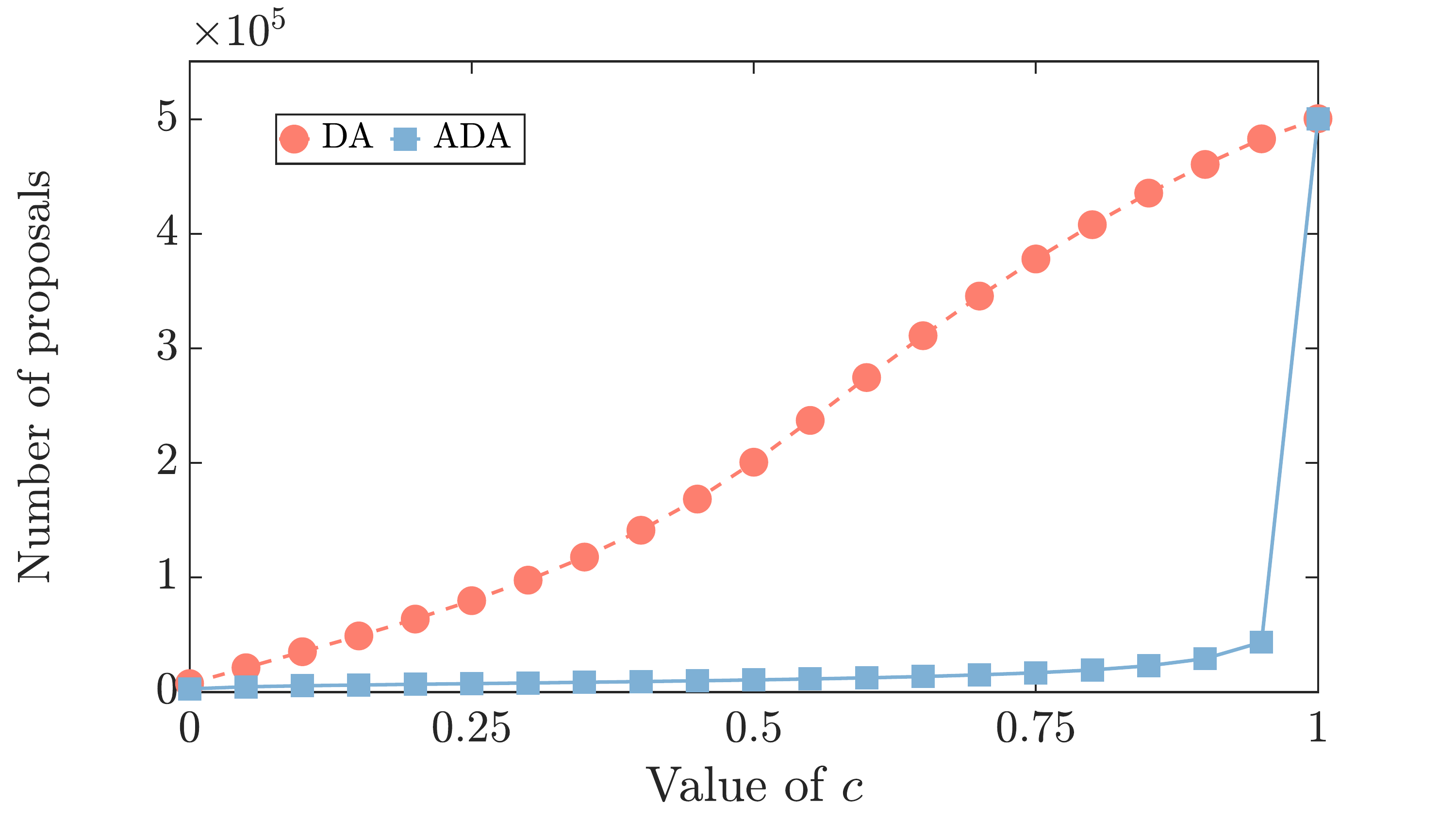}
  \caption{Number of proposals as a function of $c$.}
  \label{fig:props_by_c}
\end{figure}

Figures~\ref{fig:props_by_n} and~\ref{fig:props_by_c} show the number of proposals for each algorithm as it changes for $n$ and $c$, respectively.

Unsurprisingly, the number of proposals by both algorithms increases in $n$.
However, for DA, the number of proposals dwarfs that of ADA\@.
To see this, consider the largest instance size, $n=4,,096$, with $c=0.9$.
DA required over 7,637,702 proposals whereas ADA needed only 208,585, a reduction of 96\%.

Note that, unlike for number of rounds, the number of proposals monotonically increases with $c$.
To understand the shape, it is useful to consider why they are exactly equal when $c=1$.
In this case, for both algorithms, all men propose to the same woman in round 1.
She accepts her favourite.
In round 2, all remaining $n-1$ men propose to the agreed-upon second most desirable woman, who accepts her favourite.
And so on.
Continuing like this, it can be checked that both algorithms require $n(n+1)/2$ proposals to terminate.
When $c$ is close to but not equal to 1, DA proceeds in a manner similar to the above (but with slightly fewer concurrent proposals because some heterogeneity in preferences is present).
However, even a small amount of heterogeneity in preferences can result in a lot of pre-emptive rejections for ADA which substantially reduces the number of proposals.

We have checked the correlations for the numbers of rounds of DA and ADA, replicating the experiments at the end of Section~\ref{ssec:rounds}.
Our observations were similar except that the distributions for the number of proposals are in general narrower.

\subsection{Proportion of final pairs matched by round}
\label{ssec:final_pairs}

\begin{figure}[htbp]
  \centering
  \includegraphics[width=0.85\textwidth,trim=0.5cm 0.6cm 0.5cm 1cm,clip]{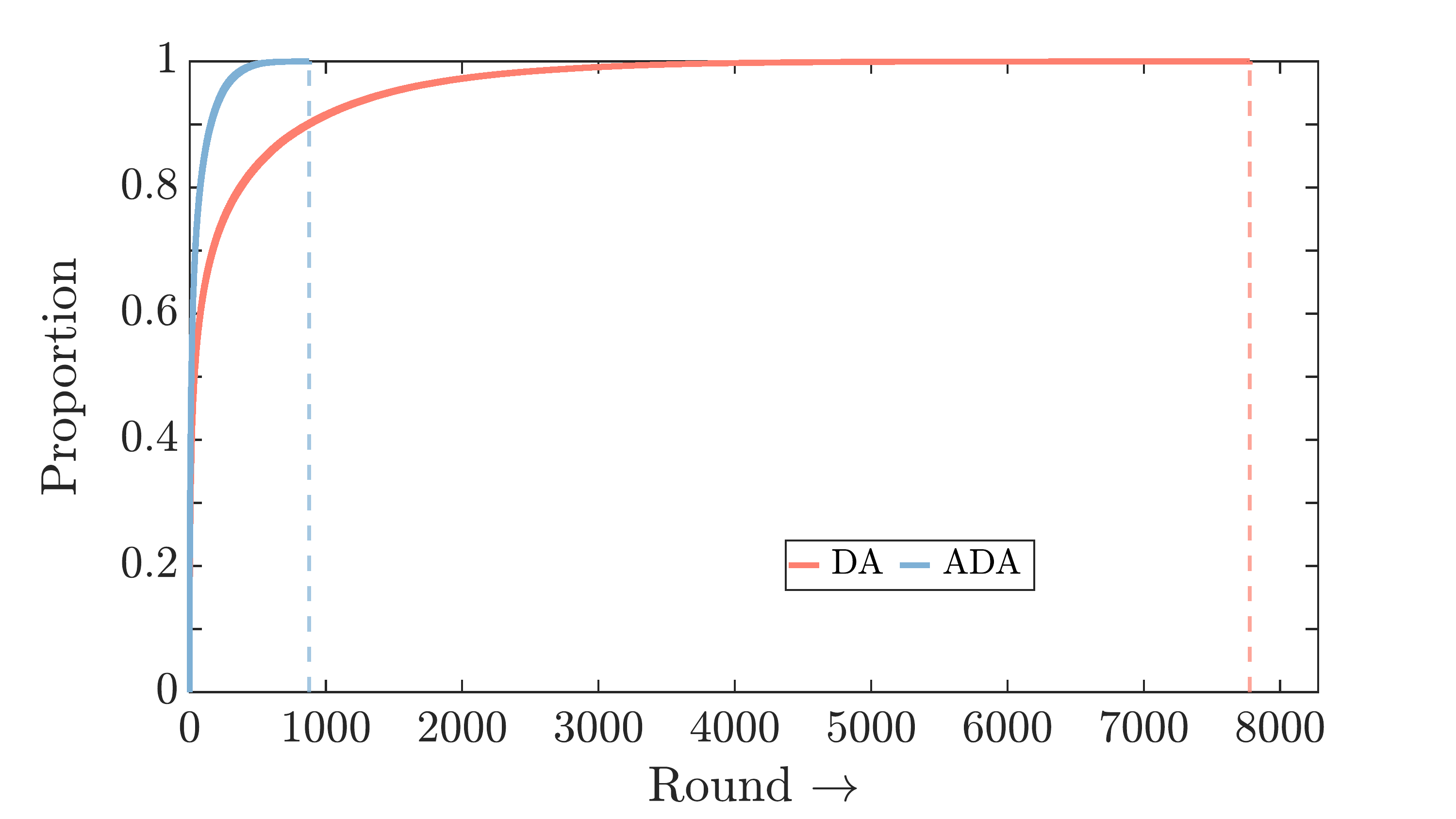}
    \caption{Proportion of final pairs matched by round for $c=0$.}
      \label{fig:cdf_c0}
\end{figure}

Another metric that might be interesting in some markets is the proportion of final pairs tentatively matched by round.
To see how the measure is computed, consider an instance of size $n=2$ where both men have identical preferences.
One of the final pairs meet in the first round, while the other pair do not meet until round 2.
In this example, our metric describes the evolution of the algorithm via a non-decreasing step function equal to 0.5 after the first round and equal to 1 when all final pairs have met.
Figure~\ref{fig:cdf_c0} below plots ``proportion of final pairs matched by round'' for both algorithms for $c=0$.

From Figure~\ref{fig:cdf_c0}, ADA arrives at the stable matching after 877 rounds while DA does not conclude until round $7,778$.
After 877 rounds, DA has matched 90\% of final pairs.
While close to completion on this measure, only 11\% of rounds are finished.
This can be contrasted with Figure~\ref{fig:probability} in Section~\ref{sec:intro} that plotted the same measure for $c=0.9$.
There, upon termination of ADA, DA had completed 7\% of rounds but had only matched 13\% of final pairs.

Despite the differences in rate of progression of the various algorithms across combinations of parameters, the shape of all four step functions plotted in both figures is similar.
By definition all functions must be weakly increasing, but we note that all are concave.
The concavity was not at all obvious to us in advance.

\subsection{Execution time}
\label{ssec:running_time}

Here we compare the running times of DA and ADA\@.
Unfortunately, due to the lack of an industry-standard implementation of DA, we had to compare ADA to our own implementation of DA\@.
Both DA and ADA are implemented in Python and include only trivial optimisations.
In fact, since the difference between the two algorithms is relatively small, the implementations are also quite similar.



\begin{figure}[htbp]
  \centering
  \includegraphics[width=0.84\textwidth,trim=0.5cm 0.6cm 0.5cm 1cm,clip]{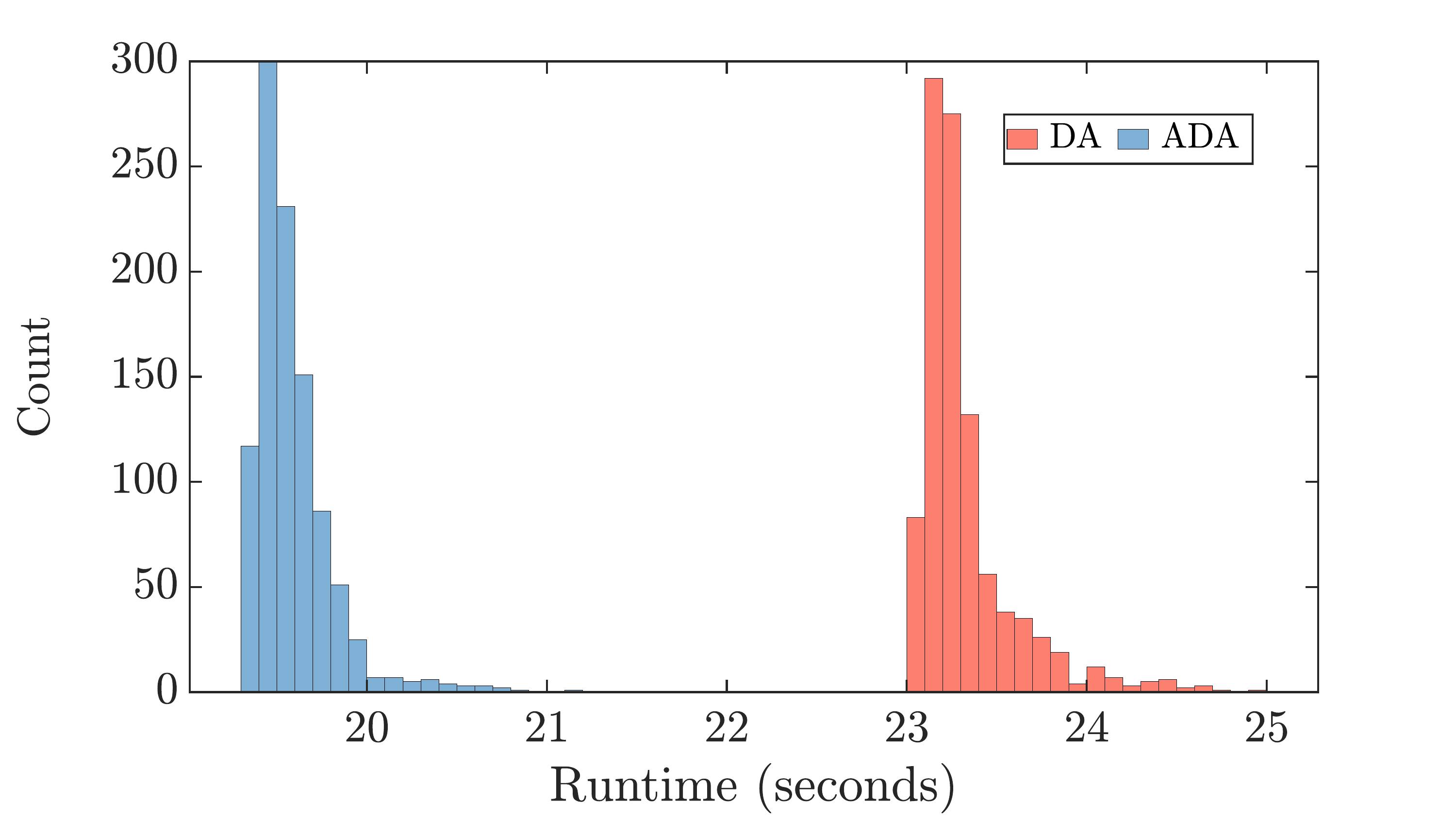}
  \caption{Runtime distributions with $n = 4,096$ and $c=0.9$.}
  \label{fig:runtime_dist}
\end{figure}

Figure~\ref{fig:runtime_dist} plots a histogram of the runtime for both algorithms for the largest markets that we considered, $n=4,096$, with $c=0.9$.
For this value of $c$, there is no overlap in the support of the two distributions.
In particular, the worst case run time for ADA is 21.18 seconds, which is shorter than the best case for DA which is 23.00 seconds.

\begin{figure}[htbp]
  \centering
  \includegraphics[width=0.84\textwidth,trim=0.5cm 0.6cm 0.5cm 1cm,clip]{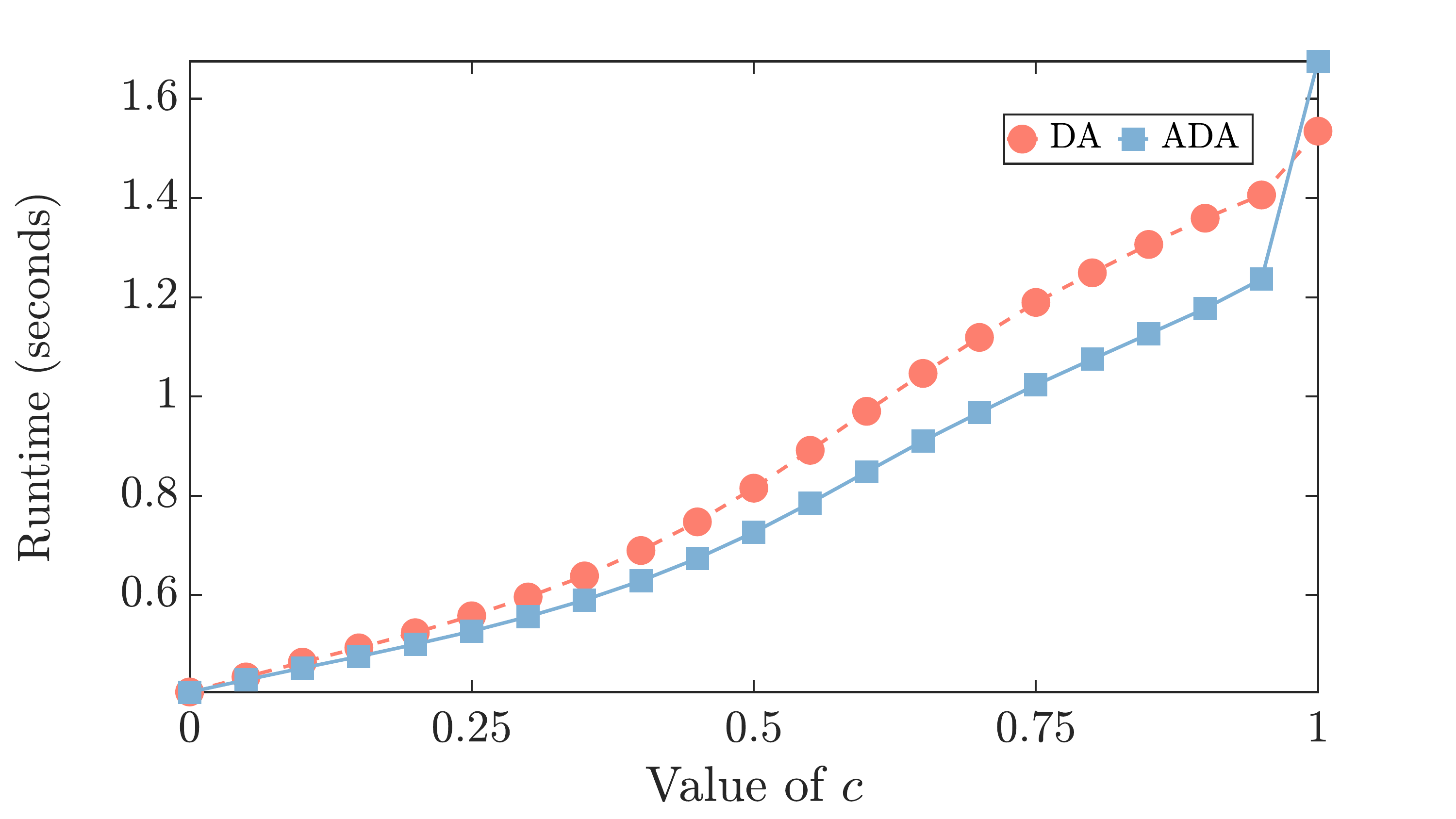}
  \caption{Runtime by $c$ with $n=1\,000$.}
  \label{fig:runtime_by_c}
\end{figure}

Figure~\ref{fig:runtime_by_c} then plots runtime as a function of $c$.
Other than the extreme point, $c=1$, in our experiments ADA is always at least as fast as DA.
For both algorithms, the average runtime is strictly increasing in $c$.
Moreover, the time saved by running ADA is increasing in $c$.
While both algorithms have almost identical execution time at $c=0$, at $c=0.95$ ADA is 15\% faster than DA.

\section{Conclusion and extensions for future work}\label{sec:conclusion}

\subsection*{Conclusion}

The deferred acceptance (DA) algorithm of \citeauthor{GaleShapley:1962:AMM} makes two outstanding contributions to the theory of matching.
It confirms the existence of a stable matching for every two-sided market and it finds the optimal stable matching from the perspective of one side.
The first property is fundamental.
The latter is of genuine practical importance because often, like when matching students to schools or doctors to hospitals, centralised matching protocols lean toward assignments that favour one side.

In this paper we proposed the accelerated deferred acceptance algorithm (ADA).
This procedure differs from DA in one minor but important way:\ it rules out proposals that will knowingly never be accepted.
This amendment does not change the output but it does generate efficiencies. 
ADA requires fewer proposals, terminates in fewer rounds, and final pairings find each other earlier.
In theory the efficiency gains are only weak.
Our computational experiments on simulated matching markets suggest that not only can these efficiency gains be strict but that they often are.

\subsection*{Extensions}

We see two natural directions in which extensions are possible.
The first involves investigating how ADA compares with DA in more general matching environments.
The second entails including the ``don't allow sure-to-be rejected proposals'' insight to other variants of the DA procedure.

\subsubsection*{1. Evaluating ADA and DA in richer environments:}
The matching environment that we have considered in this paper, Definition~\ref{def:instance}, is classified by the following three features.
\begin{enumerate}[label=(\roman*)]
  \item\label{feature:one-to-one}
        {\it a one-to-one market}:\ each participant can only match with one other.

  \item\label{feature:complete}
        {\it all preference lists are complete}:\ no participant views anyone on the other side of the market as unacceptable.

  \item\label{feature:balanced}
        {\it a balanced market}: there are the same number of men as women.

\end{enumerate}

The above features are far from universal to all matching markets.
In fact, what we have considered is somewhat artificial as rarely will all three features hold in the real world.
We held these market characteristics fixed for two reasons:\ (i) it seemed the natural starting point, and (ii) in order to keep the paper short (because there are so many ways that the above features can be relaxed).

We believe that interesting possibilities exist.
Consider for example relaxing only feature~\ref{feature:balanced} above.
That is, imagine a one-to-one matching market where preference lists are complete (features \ref{feature:one-to-one} and \ref{feature:complete} still hold) but there are more men than women.
Specifically, let $n_\Men$ denote the number of men and $n_\Wom$ the number of women and assume that $n_\Men > n_\Wom$.
When men propose, the DA algorithm will run for at least $n_\Wom$ rounds.
This must be the case because all men that end up unmatched (and by definition there is some such man) propose until they reach the end of their preference list.
But the same is not true for accelerated deferred acceptance.

\subsubsection*{2. Accelerating other variants of deferred acceptance:}

The key to our accelerated deferred acceptance algorithm is that it identifies proposals that will never be accepted because the proposee is currently matched to someone they prefer more.
Mechanically, proposals of this nature are ruled out by truncating preference lists in response to top proposals, an insight that is borrowed from the IDUA procedure \citep{BalinskiRatier:1997:,GutinNeary:2023:GEB}.
We hope that this truncation operation might prove useful to existing variants of DA.


\newpage

\appendix

\section*{APPENDIX}\label{APP}


\section{Proofs omitted from the main text}\label{app:proofs}

\begin{customthm}{1}
The man-proposing accelerated deferred acceptance returns the same matching as the man-proposing deferred acceptance algorithm.
\end{customthm}

\begin{proof}
To prove the theorem, it is sufficient to show that accelerated deferred acceptance always returns the man-optimal stable matching.

  To see that ADA must terminate, observe that at least one new proposal occurs in every round (note that $n$ proposals occur in the first round).
  Since each man has $n$ women on his preference list, the total number of proposals is bounded above by $n^2$.
  With at least one new proposal in each round and a finite number of rounds, ADA must terminate.

  To show that ADA produces a matching, suppose to the contrary.
  Then there is at least one man, call him $\man{}$, and at least one woman, $\wom{}$, who are unmatched.
  By Observation~\ref{obs:decrease}, in some round either $\man{}$ proposed to $\wom{}$ and was rejected or $\wom{}$ pre-emptively rejected $\man{}$.
But $\wom{}$ would only reject $\man{}$ if she already had a partner, and by Observation~\ref{obs:never} she never did.

To show that the matching returned by accelerated deferred acceptance is stable, again let us suppose to the contrary.
  Specifically, let the matching returned by ADA be denoted by $\match$ and suppose that $\match$ is not stable.
  Since $\match$ is unstable there must exist at least one blocking pair; denote this pair by $(\man{}, \wom{}')$.
  Since the pair $(\man{}, \wom{}')$ is blocking it must be that $\man{}$ prefers $\wom{}'$ over his match $\match(\man{})$, which means that $\man{}$ was rejected by $\wom{}$ at some round.
  But woman $\wom{}$ only rejected $\man{}$ in that round if, at that time, she had a partner that she preferred more than $\man{}$.
  Call this man $\man{}'$.
  Since women only ever trade up by Oservation~\ref{obs:never} and pairs are only ever broken by women, it must be that $\wom{}$'s final partner is at least as desirable to $\wom{}$ as $\man{}'$.
  Since preferences are strict, we have that $\match(\wom{}) \pref{\wom{}} \man{}$ and so the pair $(\man{}, \wom{}')$ cannot be a  blocking pair for $\match$.

  To complete the proof, let us now show that the stable matching $\match$ returned by accelerated deferred acceptance is the man-optimal stable matching $\match_{\Men}$ as defined in Lemma~\ref{lemma:extremal}.
  Once again suppose not.
  Then there must be some man, call him $\man{}$, who was rejected by $\wom{} = \tau(\man{})$ either in response to a proposal or preemptively.
  Consider the round where the rejection was handed out:\ by Observation~\ref{obs:never} it must be that woman $\wom{} = \tau(\man{})$ was in possession of a proposal from some man, $\man{}'$ who she prefers to $\man{}$, i.e., $\man{}' \pref{\wom{}} \man{}$.
  Since every man has a different favourite woman in the normal form by Lemma~\ref{lemma:extremal}, we must have that $\tau(\man{}') \neq \tau(\man{})$.
  Moreover, it cannot be that $\man{}'$ has proposed to his favourite woman $\tau(\man{}')$ and since $\man{}'$ is currently paired with $\wom{}$, by Observation~\ref{obs:decrease} it must be that $\wom{} \pref{\man{}'} \tau(\man{}')$.
  Now consider the man-optimal stable matching.
  By definition this must include the pairs $\big(\man{}, \tau(\man{})\big)$ and $\big(\man{}', \tau(\man{}')\big)$.
  But we have just shown that (i) $\wom{} \pref{\man{}'} \tau(\man{}')$ and (ii) $\man{}' \pref{\wom{}} \man{}$ which means that $(\man{}', \wom{})$ is a blocking pair for $\match$. This is a contradiction to $\match$ being stable.
\end{proof}

\medskip

\begin{customthm}{3}
Accelerated deferred acceptance never requires more proposals than DA.
\end{customthm}

\begin{proof}
Fix an instance of the stable matching problem and let $p_D$ denote the number of proposals when running DA and let $p_A$ denote the number of proposals when running ADA.

Let $\match_{\Men} = \set{\big(\man{1}, \tau(\man{1})\big), \dots, \big(\man{n}, \tau(\man{n})\big)}$ be the man-optimal stable matching returned by both algorithms (recall that by Theorem~\ref{thm:ADAstable}, both algorithms return $\match_{\Men}$).
By Observation~\ref{obs:decrease}, when DA terminates, each man $\man{}$ has proposed to all women strictly preferable to $\tau(\man{})$ and also to woman $\tau(\man{})$.
For ADA, no man $\man{}$ has proposed to any woman strictly worse than $\tau(\man{})$, since $\man{}$ would only propose to such a woman in the event that he was rejected by $\tau(\man{})$.
However, under ADA each man $\man{}$ may not have proposed to all women strictly preferred to  $\tau(\man{})$.
Therefore, it must be that $p_A \le p_D$.
\end{proof}


\medskip

\begin{customthm}{5}
Each proposal that is made when running ADA takes place in an earlier round than for DA.
\end{customthm}

The proof of Theorem~\ref{thm:ADAearlier} requires some terminology and notation. 
Say that $(\man{}, \wom{})$ \emph{is a pair} in matching problem $P$ if $\wom{}$ appears on the preference list of $\man{}$ and $\man{}$ appears on the preference list of $\wom{}$ in $P$.
Given an instance of the matching problem, $P$, we now define a sub-instance.

\begin{definition}
Say that matching problem $P$ is a sub-instance of matching problem $P'$ when the following properties hold.
\begin{enumerate}
\item
if $(\man{},\wom{})$ is a pair in $P$, then it is also a pair in $P'$.
\item
if man $\man{}$ prefers $\wom{}$ over $\wom{}'$ in $P$, then $\man{}$ also prefers $\wom{}$ over $\wom{}'$ in $P'$.

\item
if woman $\wom{}$ prefers $\man{}$ over $\man{}'$ in $P$, then $\wom{}$ also $\man{}$ over $\man{}'$ in $P'$.
\end{enumerate}
\end{definition}

We interpret both ADA and DA as deletion operations that generate sub-instances by deleting pairs.
We are now ready to prove Theorem~\ref{thm:ADAearlier}.

\begin{proof}
Let $P$ be an instance of the stable matchign problem.
Let $A_k(P)$ denote the sub-instance obtained after $k$ iterations of ADA and let $D_k(P)$ denote the sub-instance obtained after $k$ iterations of DA.
We will show by induction that $A_k(P)$ is a sub-instance $D_k(P)$ (denoted by $A_k(P) \subseteq D_k(P)$) for all $k=0,1,2,\ldots$
By definition, $A_0(P)=D_0(P) = P$, and clearly $A_1(P)\subseteq D_1(P)$.
Assume that $k\ge 2$, and, by the inductive hypothesis, assume that $A_r(P) \subseteq D_r(P)$ for all $r<k$.

Towards a contradiction, assume that $A_k(P)$ is not a sub-instance of $D_k(P)$.
This implies that some pair $(\man{},\wom{})$ is a pair in $A_k(P)$ but not in $D_k(P)$.
As $A_k(P) \subseteq A_{k-1}(P) \subseteq D_{k-1}(P)$, it must be that $(\man{},w)$ is a pair in $D_{k-1}(P)$.
This means that is $(\man{},\wom{})$ deleted from $D_{k-1}(P)$ by DA algorithm in iteration $k$.
Let $(\man{}',\wom{})$ be a pair such that $\man{}'$ proposes to $\wom{}$ in iteration
$k$ of DA but woman $\wom{}$ prefers $\man{}'$ over $\man{}$, which is the reason that $(\man{},\wom{})$ gets deleted from $D_{k-1}(P)$.
We now consider the following two cases separately.

\2

 {\bf Case 1, $(m',w)$ is a pair in $A_{k-1}(P)$:}
 Since $\wom{}$ is the most preferred woman for man $\man{}'$ in iteration $k$ of DA, and $A_{k-1}(P) \subseteq DA_{k-1}$ (by induction), and $(\man{}',\wom)$ is a pair in $A_{k-1}(P)$ (by assumption),  it must be that man $\man{}'$ also proposes to woman $\wom{}$ $k$'th iteration of ADA.
However, this implies that $(\man{},\wom{})$ is not a pair $A_k(P)$ because then woman $\wom{}$ prefers $\man{}'$ over $\man{}$, a contradiction to the assumption that $(\man{},\wom{})$ belongs to $A_k(P)$

\2

{\bf Case 2, $(\man{}',\wom{})$ is not a pair in $A_{k-1}(P)$:} This implies that the pair $(\man{}',\wom{})$ was deleted by ADA in the $r$'th iteration of ADA for some $r<k$.
However, since woman $\wom{}$ prefers $\man{}'$ over $\man{}$ this implies that $(\man{},\wom{})$ would also have been deleted in the $r$'th iteration (or earlier), a contradiction to the assumption that $(\man{},\wom{})$ belongs to $A_k(P)$.

\2

In both of the above cases we obtain a contradiction, implying that  $A_k(P) \subseteq D_k(P)$ for all $k=0,1,2,3\dots$, 
as desired.

\end{proof}


\newpage

\section{Solving Example~\ref{ex:bothAlgos} using both algorithms}\label{app:example}

\setcounter{example}{0}
\begin{example}\label{ex:bothAlgos}
  Consider a one-to-one two-sided matching problem with five men, $\Men = \set{\man{1}, \man{2}, \man{3}, \man{4}, \man{5}}$, and five women, $\Wom = \set{\wom{1}, \wom{2}, \wom{3}, \wom{4}, \wom{5}}$.
  The preference lists for each man and each woman are presented below, with the participants on the other side of the market listed in order of decreasing preference.
  \begin{align*}
     & \wom{1}: \, \man{5} \, , \man{4} \, , \man{1}, \,  \man{2}, \, \man{3}  \hspace{.1in}  & \blc{\blue{\man{1}}}: & \, \wom{1} \, , \wom{2} \, , \wom{3} \, , \wom{4} \, , \wom{5}  \\
     & \wom{2}: \,  \man{1} \, , \man{3} \, , \man{2}, \, \man{4}, \, \man{5}  \hspace{.1in}  & \blc{\blue{\man{2}}}: & \,\, \wom{1} \, , \wom{4} \, , \wom{5} \, , \wom{2}, \, \wom{3} \\
     & \wom{3}: \,  \man{5} \, , \man{4} \, , \man{3}, \,  \man{2}, \, \man{1}  \hspace{.1in} & \blc{\blue{\man{3}}}: & \,\, \wom{1} \, , \wom{4} \, , \wom{3} \, , \wom{5}, \, \wom{2} \\
     & \wom{4}: \,  \man{4} \, , \man{2} \, , \man{1}, \,  \man{3}, \, \man{5}  \hspace{.1in} & \blc{\blue{\man{4}}}: & \,\, \wom{4} \, , \wom{2} \, , \wom{3} \, , \wom{1}, \, \wom{5} \\
     & \wom{5}: \,  \man{5} \, , \man{1} \, , \man{3}, \,  \man{4}, \, \man{2}  \hspace{.1in} & \blc{\blue{\man{5}}}: & \,\, \wom{5} \, , \wom{4} \, , \wom{1} \, , \wom{2}, \, \wom{3} \\
  \end{align*}

  \noindent
  \extrarowheight=\aboverulesep
  \addtolength{\extrarowheight}{\belowrulesep}
  \aboverulesep=0.1pt
  \belowrulesep=0.1pt

  \begin{tabular}{L{0.5cm}C{0.5cm}C{0.5cm}||C{0.5cm}C{0.5cm}||C{0.5cm}C{0.5cm}||C{0.5cm}C{0.5cm}||C{0.5cm}C{0.5cm}}
    \multicolumn{11}{l}{Round 1:}                                                                                                                                                                                                                                                                                                                                           \\
     & \multicolumn{2}{C{1.4cm}||}{$\wom{1}$} & \multicolumn{2}{C{1.4cm}||}{$\wom{2}$} & \multicolumn{2}{C{1.4cm}||}{$\wom{3}$} & \multicolumn{2}{C{1.4cm}||}{$\wom{4}$} & \multicolumn{2}{C{1.4cm}}{$\wom{5}$}                                                                                                                                                             \\
    \cmidrule{2-11} \morecmidrules \cmidrule{2-11}
     & \cellcolor{lightgray}$\man{1}$         & \cellcolor{greengray}$\man{1}$         & \cellcolor{lightgray}                  & \cellcolor{greengray}                  & \cellcolor{lightgray}                & \cellcolor{greengray} & \cellcolor{lightgray}$\man{4}$ & \cellcolor{greengray}$\man{4}$ & \cellcolor{lightgray}$\man{5}$ & \cellcolor{greengray}$\man{5}$ \\

     & \cellcolor{lightgray}$\man{2}^*$       & \cellcolor{greengray}$\man{2}^*$       & \cellcolor{lightgray}                  & \cellcolor{greengray}                  & \cellcolor{lightgray}                & \cellcolor{greengray} & \cellcolor{lightgray}          & \cellcolor{greengray}          & \cellcolor{lightgray}          & \cellcolor{greengray}          \\

     & \cellcolor{lightgray}$\man{3}^*$       & \cellcolor{greengray}$\man{3}^*$       & \cellcolor{lightgray}                  & \cellcolor{greengray}                  & \cellcolor{lightgray}                & \cellcolor{greengray} & \cellcolor{lightgray}          & \cellcolor{greengray}          & \cellcolor{lightgray}          & \cellcolor{greengray}          \\
  \end{tabular}

  \begin{align*}
     & \wom{1}: \, \man{5} \, , \man{4} \, , \man{1}, \,  \rds{\red{\man{2}}}, \, \rds{\red{\man{3}}}  \hspace{.1in}  & \man{1}: & \, \wom{1} \, , \wom{2} \, , \wom{3} \, , \red{\wom{4}} \, , \red{\wom{5}}              \\
     & \wom{2}: \,  \man{1} \, , \man{3} \, , \man{2}, \, \man{4}, \, \man{5}  \hspace{.1in}                          & \blc{\blue{\man{2}}}: & \,\, \rds{\red{\wom{1}}} \, , \red{\wom{4}}\, , \red{\wom{5}} \, , \wom{2}, \, \wom{3}  \\
     & \wom{3}: \,  \man{5} \, , \man{4} \, , \man{3}, \,  \man{2}, \, \man{1}  \hspace{.1in}                         & \blc{\blue{\man{3}}}: & \,\, \rds{\red{\wom{1}}} \, , \red{\wom{4}} \, , \wom{3} \, , \red{\wom{5}}, \, \wom{2} \\
     & \wom{4}: \,  \man{4} \, , \red{\man{2}} \, , \red{\man{1}}, \,  \red{\man{3}}, \, \red{\man{5}}  \hspace{.1in} & \man{4}: & \,\, \wom{4} \, , \wom{2} \, , \wom{3} \, , \wom{1}, \, \red{\wom{5}}                   \\
     & \wom{5}: \,  \man{5} \, , \red{\man{1}} \, , \red{\man{3}}, \,  \red{\man{4}}, \, \red{\man{2}}  \hspace{.1in} & \man{5}: & \,\, \wom{5} \, , \red{\wom{4}} \, , \wom{1} \, , \wom{2}, \, \wom{3}                   \\
  \end{align*}
  \vspace*{-1cm}

  \begin{tabular}{L{0.5cm}C{0.5cm}C{0.5cm}||C{0.5cm}C{0.5cm}||C{0.5cm}C{0.5cm}||C{0.5cm}C{0.5cm}||C{0.5cm}C{0.5cm}}
    \multicolumn{11}{l}{Round 2:}                                                                                                                                                                                                                                                                                                                                                      \\
     & \multicolumn{2}{C{1.4cm}||}{$\wom{1}$} & \multicolumn{2}{C{1.4cm}||}{$\wom{2}$} & \multicolumn{2}{C{1.4cm}||}{$\wom{3}$} & \multicolumn{2}{C{1.4cm}||}{$\wom{4}$} & \multicolumn{2}{C{1.4cm}}{$\wom{5}$}                                                                                                                                                                        \\
    \cmidrule{2-11} \morecmidrules \cmidrule{2-11}
     & \cellcolor{lightgray}$\man{1}$         & \cellcolor{greengray}$\man{1}$         & \cellcolor{lightgray}                  & \cellcolor{greengray}$\man{2}$         & \cellcolor{lightgray}                & \cellcolor{greengray}$\man{3}$ & \cellcolor{lightgray}$\man{4}$   & \cellcolor{greengray}$\man{4}$ & \cellcolor{lightgray}$\man{5}$ & \cellcolor{greengray}$\man{5}$ \\

     & \cellcolor{lightgray}                  & \cellcolor{greengray}                  & \cellcolor{lightgray}                  & \cellcolor{greengray}                  & \cellcolor{lightgray}                & \cellcolor{greengray}          & \cellcolor{lightgray}$\man{3}^*$ & \cellcolor{greengray}          & \cellcolor{lightgray}          & \cellcolor{greengray}          \\

     & \cellcolor{lightgray}                  & \cellcolor{greengray}                  & \cellcolor{lightgray}                  & \cellcolor{greengray}                  & \cellcolor{lightgray}                & \cellcolor{greengray}          & \cellcolor{lightgray}$\man{2}^*$ & \cellcolor{greengray}          & \cellcolor{lightgray}          & \cellcolor{greengray}          \\
  \end{tabular}

  \begin{align*}
     & \wom{1}: \, \man{5} \, , \man{4} \, , \man{1}, \,  \rds{\red{\man{2}}}, \, \rds{\red{\man{3}}}  \hspace{.1in}              & \man{1}:              & \, \wom{1} \, , \wom{2} \, , \red{\wom{3}} \, , \red{\wom{4}} \, , \red{\wom{5}}                   \\
     & \wom{2}: \,  \man{1} \, , \man{3} \, , \man{2}, \, \red{\man{4}}, \, \red{\man{5}}  \hspace{.1in}                          & \blc{{\man{2}}}: & \,\, \rds{\red{\wom{1}}} \, , \rds{\red{\wom{4}}}\, , \red{\wom{5}} \, , \wom{2}, \, \red{\wom{3}} \\
     & \wom{3}: \,  \man{5} \, , \man{4} \, , \man{3}, \,  \red{\man{2}}, \, \red{\man{1}}  \hspace{.1in}                         & \blc{{\man{3}}}: & \,\, \rds{\red{\wom{1}}} \, , \rds{\red{\wom{4}}} \, , \wom{3} \, , \red{\wom{5}}, \, \wom{2}      \\
     & \wom{4}: \,  \man{4} \, , \rds{\red{\man{2}}} \, , \red{\man{1}}, \,  \rds{\red{\man{3}}}, \, \red{\man{5}}  \hspace{.1in} & \man{4}:              & \,\, \wom{4} \, , \red{\wom{2}} \, , \wom{3} \, , \wom{1}, \, \red{\wom{5}}                        \\
     & \wom{5}: \,  \man{5} \, , \red{\man{1}} \, , \red{\man{3}}, \,  \red{\man{4}}, \, \red{\man{2}}  \hspace{.1in}             & \man{5}:              & \,\, \wom{5} \, , \red{\wom{4}} \, , \wom{1} \, , \red{\wom{2}}, \, \wom{3}                        \\
  \end{align*}

  Since accelerated deferred acceptance is complete we focus only on DA from here on.
  The men that single at the beginning of Round 3 are $\man{2}$ and $\man{3}$.
  Man $\man{2}$'s top choice is $\wom{5}$ (note that $\wom{5}$ has already rejected $\man{2}$ in ADA indicated by writing \red{$\wom{5}$} instead of $\wom{5}$ in $\man{2}$'s preference list) and next on $\man{3}$'s list is the as yet unproposed to $\wom{3}$.
  So, in Round 3 of DA, $\man{2}$ proposes to and is rejected by $\wom{5}$ (as forecast by ADA) and so will be back on the market in Round 4 and $\man{3}$ proposes to $\wom{3}$ who tentatively accepts him given that she is single.

  \begin{tabular}{L{0.5cm}C{0.5cm}C{0.5cm}||C{0.5cm}C{0.5cm}||C{0.5cm}C{0.5cm}||C{0.5cm}C{0.5cm}||C{0.5cm}C{0.5cm}}
    \multicolumn{11}{l}{Round 3:}                                                                                                                                                                                                                                                                                                                                                      \\
     & \multicolumn{2}{C{1.4cm}||}{$\wom{1}$} & \multicolumn{2}{C{1.4cm}||}{$\wom{2}$} & \multicolumn{2}{C{1.4cm}||}{$\wom{3}$} & \multicolumn{2}{C{1.4cm}||}{$\wom{4}$} & \multicolumn{2}{C{1.4cm}}{$\wom{5}$}                                                                                                                                                                        \\
    \cmidrule{2-11} \morecmidrules \cmidrule{2-11}
     & \cellcolor{lightgray}$\man{1}$         & \cellcolor{greengray}$\man{1}$         & \cellcolor{lightgray}                  & \cellcolor{greengray}$\man{2}$         & \cellcolor{lightgray}$\man{3}$       & \cellcolor{greengray}$\man{3}$ & \cellcolor{lightgray}$\man{4}$ & \cellcolor{greengray}$\man{4}$ & \cellcolor{lightgray}$\man{5}$   & \cellcolor{greengray}$\man{5}$ \\

     & \cellcolor{lightgray}                  & \cellcolor{greengray}                  & \cellcolor{lightgray}                  & \cellcolor{greengray}                  & \cellcolor{lightgray}                & \cellcolor{greengray}          & \cellcolor{lightgray}          & \cellcolor{greengray}          & \cellcolor{lightgray}$\man{2}^*$ & \cellcolor{greengray}          \\
  \end{tabular}

  \begin{align}
     & \wom{1}: \, \man{5} \, , \man{4} \, , \man{1}, \,  \rds{\red{\man{2}}}, \, \rds{\red{\man{3}}}  \hspace{.1in}              & \man{1}:       & \, \wom{1} \, , \wom{2} \, , \red{\wom{3}} \, , \red{\wom{4}} \, , \red{\wom{5}}         \nonumber                \\
     & \wom{2}: \,  \man{1} \, , \man{3} \, , \man{2}, \, \red{\man{4}}, \, \red{\man{5}}  \hspace{.1in}                          & \blc{\man{2}}: & \,\, \rds{\red{\wom{1}}} \, , \rds{\red{\wom{4}}}\, , \rds{\red{\wom{5}}} \, , \wom{2}, \, \red{\wom{3}} \nonumber \\
     & \wom{3}: \,  \man{5} \, , \man{4} \, , \man{3}, \,  \red{\man{2}}, \, \red{\man{1}}  \hspace{.1in}                         & \blc{\man{3}}: & \,\, \rds{\red{\wom{1}}} \, , \rds{\red{\wom{4}}} \, , \wom{3} \, , \red{\wom{5}}, \, \wom{2}     \label{eq:prefListsFinal}       \\
     & \wom{4}: \,  \man{4} \, , \rds{\red{\man{2}}} \, , \red{\man{1}}, \,  \rds{\red{\man{3}}}, \, \red{\man{5}}  \hspace{.1in} & \man{4}:       & \,\, \wom{4} \, , \red{\wom{2}} \, , \wom{3} \, , \wom{1}, \, \red{\wom{5}}            \nonumber                  \\
     & \wom{5}: \,  \man{5} \, , \red{\man{1}} \, , \red{\man{3}}, \,  \red{\man{4}}, \, \rds{\red{\man{2}}}  \hspace{.1in}       & \man{5}:       & \,\, \wom{5} \, , \red{\wom{4}} \, , \wom{1} \, , \red{\wom{2}}, \, \wom{3}              \nonumber                
  \end{align}
  \vspace*{-1cm}
\bigskip

  The only man without a partner at the beginning of round 4 of DA is $\man{2}$, who was rejected by woman $\wom{5}$ in round 3.
  This rejection is indicated by the red box around $\man{2}$ in the preference list of $\wom{5}$ and the red box around $\wom{5}$ in the preference list of $\man{2}$.
  The top ranked woman on $\man{2}$'s preference list that he has not thus far proposed to is $\wom{2}$.
  Woman $\wom{2}$ is currently unmatched and so she (tentatively) accepts $\man{2}$.
  Every woman now has a tentative match and so the algorithm terminates and returns the collection of pairs that make up a matching.

  \begin{tabular}{L{0.5cm}C{0.5cm}C{0.5cm}||C{0.5cm}C{0.5cm}||C{0.5cm}C{0.5cm}||C{0.5cm}C{0.5cm}||C{0.5cm}C{0.5cm}}
    \multicolumn{11}{l}{Round 4:}                                                                                                                                                                                                                                                                                                                                                    \\
     & \multicolumn{2}{C{1.4cm}||}{$\wom{1}$} & \multicolumn{2}{C{1.4cm}||}{$\wom{2}$} & \multicolumn{2}{C{1.4cm}||}{$\wom{3}$} & \multicolumn{2}{C{1.4cm}||}{$\wom{4}$} & \multicolumn{2}{C{1.4cm}}{$\wom{5}$}                                                                                                                                                                      \\
    \cmidrule{2-11} \morecmidrules \cmidrule{2-11}
     & \cellcolor{lightgray}$\man{1}$         & \cellcolor{greengray}$\man{1}$         & \cellcolor{lightgray}$\man{2}$         & \cellcolor{greengray}$\man{2}$         & \cellcolor{lightgray}$\man{3}$       & \cellcolor{greengray}$\man{3}$ & \cellcolor{lightgray}$\man{4}$ & \cellcolor{greengray}$\man{4}$ & \cellcolor{lightgray}$\man{5}$ & \cellcolor{greengray}$\man{5}$ \\
  \end{tabular}
  \vspace*{0.2cm}
\end{example}

\clearpage

\bibliographystyle{plainnat}
\bibliography{accelerated.bib}

\newcommand{\noop}[1]{}
\begin{thebibliography}{25}
\providecommand{\natexlab}[1]{#1}
\providecommand{\url}[1]{\texttt{#1}}
\expandafter\ifx\csname urlstyle\endcsname\relax
  \providecommand{\doi}[1]{doi: #1}\else
  \providecommand{\doi}{doi: \begingroup \urlstyle{rm}\Url}\fi

\bibitem[Abdulkadiro{\u g}lu and
  S{\"o}nmez(2003)]{AbdulkadirogluSonmez:2003:AER}
Atila Abdulkadiro{\u g}lu and Tayfun S{\"o}nmez.
\newblock School choice: A mechanism design approach.
\newblock \emph{American Economic Review}, 93\penalty0 (3):\penalty0 729--747,
  June 2003.
\newblock \doi{10.1257/000282803322157061}.
\newblock URL
  \url{https://www.aeaweb.org/articles?id=10.1257/000282803322157061}.

\bibitem[Akbarpour et~al.(2020)Akbarpour, Li, and Gharan]{AkbarpourLi:2020:JPE}
Mohammad Akbarpour, Shengwu Li, and Shayan~Oveis Gharan.
\newblock Thickness and information in dynamic matching markets.
\newblock \emph{Journal of Political Economy}, 128\penalty0 (3):\penalty0
  783--815, 2020.
\newblock \doi{10.1086/704761}.
\newblock URL \url{https://doi.org/10.1086/704761}.

\bibitem[Balinski and Ratier(1997)]{BalinskiRatier:1997:}
Michel Balinski and Guillaume Ratier.
\newblock Of stable marriages and graphs, and strategy and polytopes.
\newblock \emph{SIAM Review}, 39\penalty0 (4):\penalty0 575--604, 1997.
\newblock \doi{10.1137/S0036144595294515}.
\newblock URL \url{https://doi.org/10.1137/S0036144595294515}.

\bibitem[B{\'o} and Hakimov(2019)]{BoHakimov:2019:EJ}
In{\'a}cio B{\'o} and Rustamdjan Hakimov.
\newblock {Iterative Versus Standard Deferred Acceptance: Experimental
  Evidence}.
\newblock \emph{The Economic Journal}, 130\penalty0 (626):\penalty0 356--392,
  07 2019.
\newblock ISSN 0013-0133.
\newblock \doi{10.1093/ej/uez036}.
\newblock URL \url{https://doi.org/10.1093/ej/uez036}.

\bibitem[B{\'o} and Hakimov(2022)]{BoHakimov:2022:GEB}
In{\'a}cio B{\'o} and Rustamdjan Hakimov.
\newblock The iterative deferred acceptance mechanism.
\newblock \emph{Games and Economic Behavior}, 135:\penalty0 411--433, 2022.
\newblock ISSN 0899-8256.
\newblock \doi{https://doi.org/10.1016/j.geb.2022.07.001}.
\newblock URL
  \url{https://www.sciencedirect.com/science/article/pii/S0899825622001129}.

\bibitem[Burdett and Coles(1997)]{BurdettColes:1997:QJE}
Ken Burdett and Melvyn~G. Coles.
\newblock Marriage and class.
\newblock \emph{The Quarterly Journal of Economics}, 112\penalty0 (1):\penalty0
  141--168, 02 1997.
\newblock ISSN 0033-5533.
\newblock \doi{10.1162/003355397555154}.
\newblock URL \url{https://doi.org/10.1162/003355397555154}.

\bibitem[Cole et~al.(1992)Cole, Mailath, and Postlewaite]{ColeMailath:1992:JPE}
Harold~L. Cole, George~J. Mailath, and Andrew Postlewaite.
\newblock Social norms, savings behavior, and growth.
\newblock \emph{Journal of Political Economy}, 100\penalty0 (6):\penalty0
  1092--1125, 1992.
\newblock ISSN 00223808, 1537534X.
\newblock URL \url{http://www.jstor.org/stable/2138828}.

\bibitem[Crawford and Knoer(1981)]{CrawfordKnoer:1981:E}
Vincent~P. Crawford and Elsie~Marie Knoer.
\newblock Job matching with heterogeneous firms and workers.
\newblock \emph{Econometrica}, 49\penalty0 (2):\penalty0 437--450, 1981.
\newblock ISSN 00129682, 14680262.
\newblock URL \url{http://www.jstor.org/stable/1913320}.

\bibitem[Doval(2022)]{Doval:2022:TE}
Laura Doval.
\newblock Dynamically stable matching.
\newblock \emph{Theoretical Economics}, 17\penalty0 (2):\penalty0 687--724,
  2022.
\newblock \doi{https://doi.org/10.3982/TE4187}.
\newblock URL \url{https://onlinelibrary.wiley.com/doi/abs/10.3982/TE4187}.

\bibitem[Dubins and Freedman(1981)]{DubinsFreedman:1981:AMM}
L.~E. Dubins and D.~A. Freedman.
\newblock Machiavelli and the gale-shapley algorithm.
\newblock \emph{The American Mathematical Monthly}, 88\penalty0 (7):\penalty0
  485--494, 1981.
\newblock ISSN 00029890, 19300972.
\newblock URL \url{http://www.jstor.org/stable/2321753}.

\bibitem[Echenique et~al.(2023)Echenique, Immorlica, and
  Vazirani]{EcheniqueImmorlica:2023:}
Federico Echenique, Nicole Immorlica, and Vijay~V. Vazirani.
\newblock \emph{Online and Matching-Based Market Design}.
\newblock Cambridge University Press, Cambridge, 2023.
\newblock ISBN 9781108831994.
\newblock \doi{DOI: 10.1017/9781108937535}.
\newblock URL
  \url{https://www.cambridge.org/core/product/604CA9FF1396C489D6497CF336368524}.

\bibitem[Eeckhout(1999)]{Eeckhout:1999:IER}
Jan Eeckhout.
\newblock Bilateral search and vertical heterogeneity.
\newblock \emph{International Economic Review}, 40\penalty0 (4):\penalty0
  869--887, 1999.
\newblock ISSN 00206598, 14682354.
\newblock URL \url{http://www.jstor.org/stable/2648705}.

\bibitem[Gale and Shapley(1962)]{GaleShapley:1962:AMM}
D.~Gale and L.~S. Shapley.
\newblock College admissions and the stability of marriage.
\newblock \emph{The American Mathematical Monthly}, 69\penalty0 (1):\penalty0
  9--15, 1962.
\newblock ISSN 00029890.
\newblock URL \url{http://www.jstor.org/stable/2312726}.

\bibitem[Gutin et~al.(2024)Gutin, Neary, and Yeo]{GutinNeary:2024:arXiv}
Gregory Gutin, Philip~R. Neary, and Anders Yeo.
\newblock Finding all stable matchings with assignment constraints, 2024.
\newblock URL \url{https://arxiv.org/abs/2204.03989}.

\bibitem[Gutin et~al.(2023)Gutin, Neary, and Yeo]{GutinNeary:2023:GEB}
Gregory~Z. Gutin, Philip~R. Neary, and Anders Yeo.
\newblock Unique stable matchings.
\newblock \emph{Games and Economic Behavior}, 141:\penalty0 529--547, 2023.
\newblock ISSN 0899-8256.
\newblock \doi{https://doi.org/10.1016/j.geb.2023.07.010}.
\newblock URL
  \url{https://www.sciencedirect.com/science/article/pii/S0899825623001033}.

\bibitem[Hayek(1945)]{Hayek:1945:AER}
F.~A. Hayek.
\newblock The use of knowledge in society.
\newblock \emph{American Economic Review}, 1945.

\bibitem[Holzman and Samet(2014)]{HolzmanSamet:2014:GEB}
Ron Holzman and Dov Samet.
\newblock Matching of like rank and the size of the core in the marriage
  problem.
\newblock \emph{Games and Economic Behavior}, 88:\penalty0 277--285, 2014.
\newblock ISSN 0899-8256.
\newblock \doi{https://doi.org/10.1016/j.geb.2014.10.003}.
\newblock URL
  \url{https://www.sciencedirect.com/science/article/pii/S0899825614001468}.

\bibitem[Kelso and Crawford(1982)]{KelsoCrawford:1982:E}
Jr~Kelso, Alexander~S and Vincent~P Crawford.
\newblock Job matching, coalition formation, and gross substitutes.
\newblock \emph{Econometrica}, 50\penalty0 (6):\penalty0 1483--1504, November
  1982.
\newblock URL
  \url{http://ideas.repec.org/a/ecm/emetrp/v50y1982i6p1483-1504.html}.

\bibitem[Milgrom and Segal(2020)]{MilgromSegal:2020:JPE}
Paul Milgrom and Ilya Segal.
\newblock Clock auctions and radio spectrum reallocation.
\newblock \emph{Journal of Political Economy}, 128\penalty0 (1):\penalty0
  1--31, 2020.
\newblock \doi{10.1086/704074}.
\newblock URL \url{https://doi.org/10.1086/704074}.

\bibitem[Nagel(1995)]{Nagel:1995:AER}
Rosemarie Nagel.
\newblock Unraveling in guessing games: An experimental study.
\newblock \emph{American Economic Review}, 85\penalty0 (5):\penalty0 1313--26,
  December 1995.
\newblock URL
  \url{http://ideas.repec.org/a/aea/aecrev/v85y1995i5p1313-26.html}.

\bibitem[Nisan et~al.(2007)Nisan, Roughgarden, Tardos, and
  Vazirani]{NisanRoughgarden:2007:}
Noam Nisan, Tim Roughgarden, \'Eva Tardos, and Vijay~V. Vazirani.
\newblock \emph{Algorithmic Game Theory}.
\newblock Cambridge University Press, New York, NY, USA, 2007.

\bibitem[Roth(1982)]{Roth:1982:MOR}
Alvin~E. Roth.
\newblock The economics of matching: Stability and incentives.
\newblock \emph{Mathematics of Operations Research}, 7\penalty0 (4):\penalty0
  617--628, 1982.
\newblock \doi{10.1287/moor.7.4.617}.
\newblock URL \url{https://doi.org/10.1287/moor.7.4.617}.

\bibitem[Roth(2008)]{Roth:2008:IJGT}
Alvin~E. Roth.
\newblock Deferred acceptance algorithms: history, theory, practice, and open
  questions.
\newblock \emph{International Journal of Game Theory}, 36\penalty0
  (3):\penalty0 537--569, 2008.

\bibitem[Stahl and Wilson(1994)]{StahlWilson:1994:JEBO}
Dale~O. Stahl and Paul~W. Wilson.
\newblock Experimental evidence on players' models of other players.
\newblock \emph{Journal of Economic Behavior and Organization}, 25\penalty0
  (3):\penalty0 309 -- 327, 1994.
\newblock ISSN 0167-2681.
\newblock \doi{10.1016/0167-2681(94)90103-1}.
\newblock URL
  \url{http://www.sciencedirect.com/science/article/pii/0167268194901031}.

\bibitem[Stahl and Wilson(1995)]{StahlWilson:1995:GEB}
Dale~O. Stahl and Paul~W. Wilson.
\newblock On players' models of other players: Theory and experimental
  evidence.
\newblock \emph{Games and Economic Behavior}, 10\penalty0 (1):\penalty0 218 --
  254, 1995.
\newblock ISSN 0899-8256.
\newblock \doi{10.1006/game.1995.1031}.
\newblock URL
  \url{http://www.sciencedirect.com/science/article/pii/S0899825685710317}.

\end{thebibliography}

\end{document}